\documentclass[epj]{svjour}
\usepackage{amssymb}
\usepackage{amsmath}
\usepackage{array} 
\usepackage{epsfig}
\usepackage{graphics}
\usepackage{colortbl}
\usepackage{hhline}
\usepackage{multirow}
\begin{document}


%
\title{Single and double diffractive dissociation and\\ 
the problem of extraction of the proton-Pomeron cross-section.}
\author{V.A.~Petrov\thanks{\emph{e-mail:} Vladimir.Petrov@cern.ch}\inst{1},
R.A.~Ryutin\thanks{\emph{e-mail:} Roman.Rioutine@cern.ch}\inst{1}
}                     
%
%
\institute{{\small Institute for High Energy Physics, NRC ``Kurchatov Institute'', Protvino {\it 142 281}, Russia}}
%
%
\abstract{Diffractive dissociation processes are analysed in the 
framework of covariant reggeization. We have considered 
the general form of hadronic tensor
and its asymptotic behaviour for $t\rightarrow 0$ in the case of
conserved tensor currents before reggeization. Resulting expressions
for differential cross-section
of single dissociation (SD) process ($pp \rightarrow p M$),
double dissociation (DD) ($pp\rightarrow M_1 M_2$)
and for the proton-Pomeron cross-section are given 
in detail, and corresponding problems of the approach are discussed.
\PACS{
     {11.55.Jy}{Regge formalism}   \and
      {12.40.Nn}{Regge theory, duality, absorptive/optical models} \and
      {13.85.Ni}{Inclusive production with identified hadrons}
     } 
} 
\authorrunning{Single and double diffractive dissociation and}
\titlerunning{the problem of extraction of the proton-Pomeron cross-section}
\maketitle
%

\section*{Introduction}

Reaching the new energy frontier at the LHC opens new opportunities for further study 
of the elastic and inelastic diffractive processes. The latter particularly 
need more data which up to now remained insufficient for a thorough theoretical analysis.  
There is a standard set of the processes which are related to 
the same driving interaction agent, the Pomeron  and  Reggeon exchanges: 

\begin{itemize}
\item Elastic scattering (ES)
\item Single Diffractive Dissociation (SD)
\item Double Diffractive Dissociation (DD)
\item Exclusive Diffractive  Central  Production (EDCP)
\end{itemize}

The study of these processes is aimed  to explore the properties of Pomerons:
\begin{itemize}
\item Pomeron coupling to the protons (ES \& SD)
\item Pomeron-proton interaction  (SD \& DD)
\item Pomeron-Pomeron interactions (EDCP)
\item Pomeron quark-gluon structure (jet- and high-mass production in SD and EDCP)
\item Pomeron trajectory (all)
\end{itemize}

It is extremely important to try to extract from the data such fundamental characteristics as 
\begin{itemize}
\item Pomeron-proton total cross-section
\item Cross-sections of  Pomeron-Pomeron to exclusive hadron states (say, pion-pion)
\item Pomeron-Pomeron to two-jets. 
\end{itemize}

It is clear that it would be very welcomed to do that in a way which is model independent 
as much as possible. One, however, has to note that the very notion of the Pomeron is far from 
being well defined and established, even in theory. Perennial attempts to find the Pomeron 
trajectory (or at least its intercept and slope) in the framework of QCD had little 
success up to now. In various models not only the Pomeron has different properties, but 
there are options with two, three  or even infinitely many 
different Pomerons. In such a situation it is almost impossible to avoid a 
significant model dependence of  the characteristics enlisted above. We, nonetheless, believe 
that  ``If you can't get a horse, ride a cow''. 

The attempts to extract the Po\-me\-ron-proton and 
Po\-me\-ron-Pomeron cross-sections from the data on SD and DD  were 
undertaken quite a long ago (see e.g., \cite{0,1}). The values of 
the cross-sections appeared very small (less than 6~mb) in compare, say, with 
pion-proton cross-sections. It seems strange because the pomeron, as we 
believe, consists mostly of strong-interacting  gluon fields and at small 
transfers its ``mass'' squared ($\sim t$) lies not
very far from the pion mass squared ($\sim 0.02\,\mathrm{GeV}^2$). From the 
latter viewpoint this smallness seems strange. Usually it is being related 
with the smallness of the 3-pomeron vertex at small momenta. Up to now the 
reason of this smallness is not clear. In what follows 
we will address these problems in the  framework of the chosen approach .

\section*{Hadronic tensor}
\subsection*{Kinematics}

Let us consider first the general process\\
$p(k)+p(p)\to M_1(k^{\prime})+M_2(p^{\prime})$,\\
where\\
$p^2=k^2=m^2$, 
${k^{\prime}}^2=(k-q)^2=M_1^2$, ${p^{\prime}}^2=(p+q)^2=M_2^2$, 
$q^2=-Q^2\equiv t$, $s=(p+k)^2$,  
and define its kinematic
quantities. In the center of mass frame we have 
(for any momenta we use the notation
$p\equiv (p_0,p_1,p_2,p_3)$)

\begin{eqnarray}
p&=&\left( \frac{\sqrt{s}}{2},\;0,\;0,\;\beta\frac{\sqrt{s}}{2}\right),\;
k=\left( \frac{\sqrt{s}}{2},\;0,\;0,\;-\beta\frac{\sqrt{s}}{2}\right)\nonumber\\
q&=&(q_0,\;q_1,\;q_2,\;q_3),\; q_0=\frac{kq+pq}{\sqrt{s}},\; q_3=\frac{kq-pq}{\beta\sqrt{s}},\nonumber\\
(q_1,q_2)&=&\rho_q\sin\theta(\cos\phi,\sin\phi),\;\rho_q^2=q_0^2+Q^2,\nonumber
\end{eqnarray}
\begin{eqnarray}
\sin^2\theta&=&\left(
Q^2-\frac{m^2(pq^2+kq^2)}{pk^4-m^4}
\right)/\left(
Q^2+\frac{(pq+kq)^2}{2(pk+m^2)}
\right),\nonumber\\
\beta&=&\sqrt{1-\frac{4m^2}{s}},\nonumber\\
kq&=&\frac{m^2-Q^2-M_1^2}{2},\; pq=\frac{-m^2+Q^2+M_2^2}{2},
\nonumber
\\
Q_{\mathrm{min}}^2&=&
\frac{m^2((kq)^2+(pq)^2)-2\,pk\,kq\,pq}{(pk)^2-m^4}
\nonumber\\
&=& \frac{s}{2}
\left(\phantom{\sqrt{
\left( 1-4\xi_m\right)
\left( \left( 1-\xi_1-\xi_2\right)^2-4\xi_1\xi_2\right)}}\hspace*{-5.7cm}
1-\xi_1-\xi_2-2\xi_m-\right.\nonumber\\
&&\left.\sqrt{
\left( 1-4\xi_m\right)
\left( \left( 1-\xi_1-\xi_2\right)^2-4\xi_1\xi_2\right)}
\right),\nonumber\\
\xi_i&=&M_i^2/s,\; \xi_m=m^2/s.\label{eq:genkin}
\end{eqnarray}

\begin{figure}[ht!]  
 \includegraphics[width=0.49\textwidth]{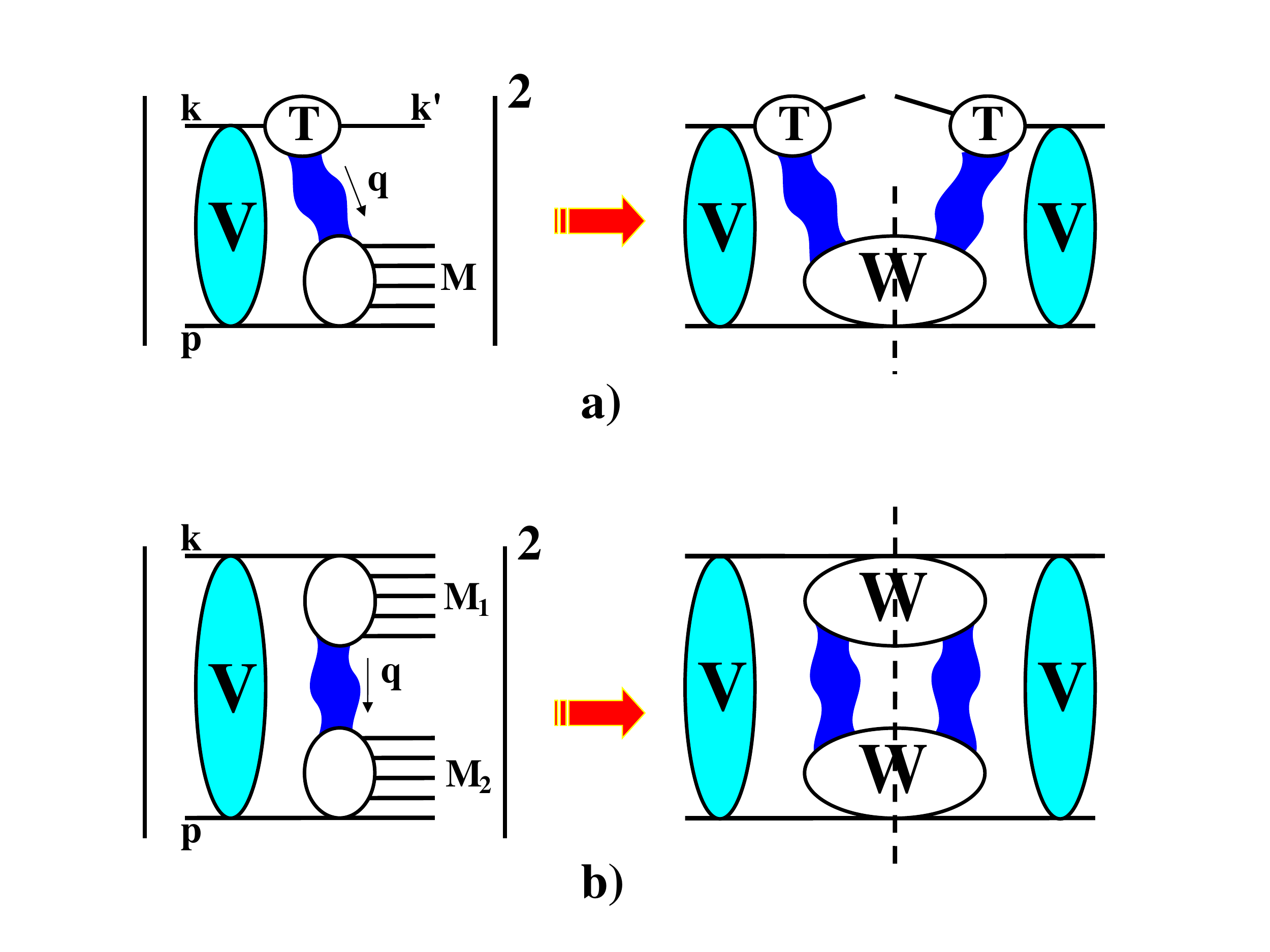} 
  \caption{\label{fig1:processes} Scheme of calculation of the full SD (a) and DD (b) cross-sections. Wavy fat lines design Reggeons.}
\end{figure}

In the case of SD (Fig.~\ref{fig1:processes}(a)) we can write
\begin{eqnarray}
&& M_2=M,\; M_1=m,\; \xi_1=\xi_m,\; \xi_2=\xi,\; Q_{\mathrm{min}}\simeq m\xi,\;\nonumber\\
&& Q_{\mathrm{min}}\ll Q\ll m,\; m\ll M\ll \sqrt{s}.\label{eq:SDlims}
\end{eqnarray}
For DD (Fig.~\ref{fig1:processes}(b)) we have
\begin{eqnarray}
&& \xi_{1,2}\gg\xi_m,\; Q_{\mathrm{min}}\simeq \sqrt{s\xi_1\xi_2},\;\nonumber\\
&& Q_{\mathrm{min}}\ll Q\ll m,\; m\ll M_{1,2}\ll \sqrt{s}.\label{eq:DDlims}
\end{eqnarray}

\subsection*{Covariant reggeization}

In order to calculate SD and DD cross-sections 
we use the amplitudes with meson exchanges of arbitrary 
spins with subsequent reggeization. Basic elements of 
such an approach are vertex functions 
$T^{\mu_1\cdots\mu_{J}}(k,q)$, $T^{\nu_1\cdots\nu_{J^{\prime}}}(k,q)$, where
\begin{equation}
 \label{eq:TvertexDef}
 T^{\mu_1\dots\mu_J}(k,q)=<k-q| I^{\mu_1\dots\mu_J}|p>,
\end{equation}
hadronic tensor
\begin{eqnarray}
&& W^{\mu_1...\mu_J\nu_1...\nu_{J^{\prime}}}(p,q)=\nonumber\\
&&\int d^4x\; e^{iqx}
\left< p|\; I^{\mu_1...\mu_J}(x)I^{\nu_1...\nu_{J^{\prime}}}(0)\; |p \right>\quad{,}
\label{eq:WtensorDef}
\end{eqnarray}
and propagators $d(J,t)/(m^2(J)-t)$ which have the poles at
\begin{equation}
\label{eq:polesJ}
m^2(J)-t=0,\; \mbox{i.e.}\; J=\alpha_{{\mathbb P}}(t)\;,
\end{equation}
after an appropriate analytic continuation of the signatured amplitudes 
in $J$. We assume that this pole, where $\alpha_{\mathbb P}$ is the Pomeron 
trajectory, gives, by definition,  the dominant contribution at high energies 
after having taken the corresponding residues. At this stage
we do not take into account absorptive corrections denoted by $V$ blobes
in Fig.~\ref{fig1:processes}, we will calculate them
in the next section.

$I^{\mu_1\dots\mu_{J}}$ is the current operator related to the hadronic spin-$J$ Heisenberg field operator,
\begin{equation}
\label{eq:KGJ}
\left( \square + m_J^2\right) \Phi^{\mu_1\dots\mu_J}(x)=I^{\mu_1\dots\mu_J}(x)\;,
\end{equation}
and
\begin{eqnarray}
&&\partial_{\mu}I^{\mu_1...\mu...\mu_J}=
\partial_{\nu}I^{\nu_1...\nu...\nu_{J^{\prime}}}=0\;{;}\label{eq:Icond1}\\
&&I^{\mu_1...\mu_J}=I^{\left(\mu_1...\mu_J\right)}\;{;}\;
I^{\nu_1...\nu_{J^{\prime}}}=I^{\left(\nu_1...\nu_{J^{\prime}}\right)}\;{;}
\label{eq:Icond2}\\
&&g_{\mu_i\mu_k}I^{\mu_1...\mu_i...\mu_k...\mu_J}=
g_{\nu_i\nu_k}I^{\nu_1...\nu_i...\nu_k...\nu_{J^{\prime}}}=0.\label{eq:Icond3}
\end{eqnarray}
In terms of tensor $W^{\mu_1...\mu_J\nu_1...\nu_{J^{\prime}}}$ conditions (\ref{eq:Icond1})-(\ref{eq:Icond3})
look like Rarita-Schwinger conditions for irreducible representations of the Poincar\'e algebra
\begin{eqnarray}
&&\hspace*{-0.7cm} q_{\lambda}W^{\mu_1...\lambda ...\nu_{J^{\prime}}}=0\;{;}\label{eq:Wcond1}\\
&&\hspace*{-0.7cm} W^{\mu_1...\mu_J\nu_1...\nu_{J^{\prime}}}=
W^{\left(\mu_1...\mu_J\right)\left(\nu_1...\nu_{J^{\prime}}\right)}\;{;}
\label{eq:Wcond2}\\
&&\hspace*{-0.7cm}g_{\mu_i\mu_k}W^{\mu_1...\mu_i...\mu_k...\mu_J...}=
g_{\nu_i\nu_k}W^{...\nu_1...\nu_i...\nu_k...\nu_{J^{\prime}}}=0{.}\label{eq:Wcond3}
\end{eqnarray}
Similar conditions are imposed on T-tensors
\begin{eqnarray}
&&T^{\mu_1\dots\mu_i\dots\mu_j\dots\mu_J}=T^{\mu_1\dots\mu_j\dots\mu_i\dots\mu_J}\label{eq:Tcond1}\\
&&q_{\mu_i}T^{\mu_1\dots\mu_i\dots\mu_J}=0\label{eq:Tcond2}\\
&&g_{\mu_i\mu_j}T^{\mu_1\dots\mu_i\dots\mu_j\dots\mu_J}=0.\label{eq:Tcond3}
\end{eqnarray}

Let us define main structures that we use in the paper:
\begin{eqnarray}
 G_{\alpha\beta}&=&g_{\alpha\beta}-\frac{q_{\alpha}q_{\beta}}{q^2}\;{;}\nonumber\\
 P_{\alpha}&=&\left( p_{\alpha}-q_{\alpha}\frac{pq}{q^2}\right)/
\sqrt{m^2-(pq)^2/q^2},\;P^2=1,\nonumber\\
 K_{\alpha}&=&\left( k_{\alpha}-q_{\alpha}\frac{kq}{q^2}\right)/
\sqrt{m^2-(kq)^2/q^2},\;K^2=1;\nonumber\\
 G_{\alpha\beta}P^{\beta}&=&P_{\alpha},\; G_{\alpha\beta}K^{\beta}=K_{\alpha},\nonumber\\
 g_{\alpha\beta}G^{\alpha\beta}&=&G_{\alpha\beta}G^{\alpha\beta}=3.
\label{eq:structures}
\end{eqnarray} 

For vertex
functions $T$ we can obtain the following tensor decomposition:
\begin{eqnarray}
&& T^{(J)}\equiv T^{\mu_1\dots\mu_{J_i}}(k,q)=\nonumber\\
&&\phantom{T^{(J)}\equiv}\label{eq:Ttensor0} F_J(t) 
\sum_{n=0}^{\left[\frac{J}{2}\right]} 
\frac{{\mathbb C}_J^n}{{\mathbb C}_J^0} \left( K^{(J-2n)}G^{(n)}\right)\;,\\
&&\phantom{T^{(J)}\equiv}\label{eq:CJn0}{\mathbb C}_J^n=
\frac{(-1)^n (2(J-n))\mbox{!}}{(J-n)\mbox{!}n\mbox{!}(J-2n)\mbox{!}},
\end{eqnarray}
where
tensor structures $\left( K^{(J-2n)}G^{(n)}\right)^{\mu_1\dots\mu_{J}}$ satisfy only two 
conditions~(\ref{eq:Tcond1}),(\ref{eq:Tcond2}) (transverse-symmetric)
\begin{eqnarray}
&&\left( K^{(J-2n)}G^{(n)}\right)=\nonumber\\
&&\label{KGsym} \frac{K^{\; (\mu_1}\!\!\cdot_{\dots}\!\!\cdot K^{\; \mu_{J-2n}}
G^{\mu_{J-2n+1}\mu_{J-2n+2}}\!\!\cdot_{\dots}\!\!\cdot G^{\mu_{J-1}\mu_{J})}}{N_{J}^n},\\
&& \label{coefPGsym}N_{J}^n=\frac{J\mbox{!}}{2^n n\mbox{!}(J-2n)\mbox{!}}.
\end{eqnarray}
Coefficients ${\mathbb C}_{J}^n$ in~(\ref{eq:Ttensor0}) can be obtained from 
the condition~(\ref{eq:Tcond3}) which
leads to the recurrent set of equations (see~\cite{ryutin:eddelowm}).

Now we are to calculate the hadronic tensor. Let us introduce the following  notations
for tensor constructions in use:
\begin{eqnarray}
G^{(l)}_M&=&G_{(\mu_{i_1}\mu_{i_2}}G_{\mu_{i_3}\mu_{i_4}}\cdot
...\cdot G_{\mu_{i_{l-1}}\mu_{i_l})},\nonumber\\
G^{(s)}_N&=&G_{(\nu_{i_1}\nu_{i_2}}G_{\nu_{i_3}\nu_{i_4}}\cdot
...\cdot G_{\nu_{i_{s-1}}\nu_{i_s})},\nonumber\\
 P^{(m)}_M&=&P_{\mu_{i_1}}\cdot ...\cdot P_{\mu_{i_m}},
P^{(n)}_N=P_{\nu_{i_1}}\cdot ...\cdot P_{\nu_{i_n}}\;{;}\nonumber\\
 G^{(k)}_{MN}&=& 
G_{(\mu_{i_1}(\nu_{i_1}}G_{\mu_{i_2}\nu_{i_2}}\cdot ...\cdot
G_{\mu_{i_k})\nu_{i_k})}.\label{eq:Wstr1}
\end{eqnarray}
where
$\left(\mu_{i_1}\mu_{i_2}...\right)\rightarrow M$ and
$\left(\nu_{i_1}\nu_{i_2}...\right)\rightarrow N$,
cor\-res\-pon\-ding\-ly. The generic construction 
$\left(P^{(m)}_MG^{(l)}_MG^{(k)}_{MN}G^{(s)}_NP^{(n)}_N\right)$ 
is symmetric in
each group ($M$ and $N$) of indices and trans\-ver\-sal in every index. 
For $J=J^{\prime}$ (for simplicity we consider here only this case) 
we can introduce the notation
\begin{eqnarray}
\{n_1kn_2\}&=&
\frac{
\left(
P^{(J-2n_1-k)}_MG^{(n_1)}_MG^{(k)}_{MN}G^{(n_2)}_NP^{(J-2n_2-k)}_N
\right)
}
{N_{n_1kn_2}
},\nonumber\\
 N_{n_1kn_2}&=&
\frac{\left(J\mbox{!}\right)^2}{k\mbox{!}n_1\mbox{!}n_2\mbox{!}(J-2n_1)\mbox{!}(J-2n_2)\mbox{!}2^{n_1+n_2}}\label{eq:n1kn2}
\end{eqnarray}
The hadronic tensor can be now written in the form
\begin{equation}
W_{MN}=\sum\limits_{k=0}^{J-1}\sum\limits_{n_{1,2}=0}^{\left[ \frac{J}{2}\right]}
f^k_{n_1n_2}\{n_1kn_2\}+f^J_{0,0}\{0J0\}
\label{eq:Wdecomp1}
\end{equation}
To obtain TST (transverse-symmetric-traceless) construction we have to
solve recurrent set of equations for $f^k_{n_1n_2}$ from trace\-less\-ness
conditions $g_{\mu_i\mu_j}W_{MN}=0$ and\linebreak
$g_{\nu_i\nu_j}W_{MN}=0$, or
$g_{\mu_i\mu_j}W_{MN}=0$ and $f^k_{n_1n_2}=f^k_{n_2n_1}$:
\begin{eqnarray}
&& f^k_{n_1n_2}=f^k_{n_2n_1},\; n_1\le n_2\le \left[ \frac{J-k}{2}\right],\label{eq:symcondeq}\\
&& f^k_{n_1n_2}(J-2n_1-k)(J-2n_1-k-1)+\nonumber\\
&& f^{(k+2)}_{n_1(n_2-1)}(k+1)(k+2)
\Theta(n_2-1)\times\nonumber\\
&&\phantom{f^{(k+2)}_{n_1(n_2-1)}} \Theta(\left[ \frac{J-k-2}{2}\right]+1-n_2)+\nonumber\\
&& f^{(k+1)}_{n_1n_2}(2(k+1)(J-2n_1-k-1))\times\nonumber\\
&&\phantom{f^k_{(n_1+1)n_2}}
\Theta(\left[ \frac{J-k-1}{2}\right]-n_2)+\nonumber\\
&& f^k_{(n_1+1)n_2} (2(n_1+1)(2J-2n_1-1))\times\nonumber\\
&&\phantom{f^k_{(n_1+1)n_2}}
\Theta(\left[ \frac{J-k}{2}\right]-1-n_1)=0,\nonumber\\
&& J\ge 2,\;  0\le k\le J-2,\;\nonumber\\
&& 0\le n_1\le \left[ \frac{J-k-2}{2}\right],\; 0\le n_2\le \left[ \frac{J-k}{2}\right]
\label{eq:SPMeq},
\end{eqnarray}
where
$$
\Theta (x) =\left\{ 1, x\ge 0\atop{0, x<0}\right. . 
$$
Finally it is possible to express all the functions 
in terms of $f^k_{00}$, $k=0,J$. Usually in the literature they are
denoted as structure functions of the hadronic tensor 
$W_{J-k+1}=f^k_{00}$.

Let us consider the limit $Q\to Q_{\mathrm min}\sim 0$. From the condition
of finiteness of the hadronic tensor at $ q=0 $ we can obtain in this limit  $J$
relations among the structure functions $W_i$, and, finally,
we can express the hadronic tensor in terms of $W_1$.

As an illustrative example, we consider cases $J=1,2,3$. 
Hadronic tensors look like
\begin{eqnarray}
&&\hspace*{-0.3cm} J=1,\; W_{MN}=W_1 \{010\}+W_2 \{000\},\nonumber\\
&&\hspace*{-0.3cm} J=2,\; W_{MN}=\nonumber\\
&&\hspace*{-0.3cm} W_1 \left[
\{020\}-\frac{1}{3}\{101\}
\right]+\nonumber\\
&&\hspace*{-0.3cm} W_2 \left[
\{010\}-\frac{1}{3}\left( \{001\}+\{100\}\right)+\frac{1}{9}\{101\}
\right]+\nonumber\\
&&\hspace*{-0.3cm} W_3 \left[ 
\{000\}-\frac{1}{3}\left( \{001\}+\{100\}\right)+\frac{1}{9}\{101\}
\right],\nonumber
\end{eqnarray}
\begin{eqnarray}
&&\hspace*{-0.3cm} J=3,\; W_{MN}=\nonumber\\
&&\hspace*{-0.3cm} W_1 \left[ 
\{030\}-\frac{3}{5}\{111\}
\right]+\nonumber\\
&&\hspace*{-0.3cm} W_2 \left[ 
\{020\}-\frac{2}{5}\left( \{011\}+\{110\}\right)+\frac{2}{25}\{111\}-
\frac{1}{25}\{101\}
\right]+\nonumber\\
&&\hspace*{-0.3cm} W_3 \left[ 
\{010\}-\frac{2}{5}\left( \{001\}+\{100\}\right)-
\frac{1}{5}\left( \{011\}+\{110\}\right)+\right.\nonumber\\
&&\hspace*{-0.3cm} \phantom{W_3[}
\left.\frac{1}{25}\{111\}+
\frac{8}{25}\{101\}
\right]+\nonumber\\
&&\hspace*{-0.3cm} W_4 \left[ 
\{000\}-\frac{3}{5}\left( \{001\}+\{100\}\right)+
\frac{9}{25}\{101\}
\right],\label{eq:TST123}
\end{eqnarray}
and for $Q\to Q_{\mathrm min}$
\begin{eqnarray}
&& J=1,\; W_2=-W_1,\; W_{MN}=W_1 \left[ \{010\}-\{000\}\right] ,\nonumber\\
&& J=2,\; W_2=-2W_1,\; W_3=\frac{1}{2}W_1,\;\nonumber\\
&& W_{MN}=\nonumber\\
&& W_1 \left[
\{020\}-2\{010\}+\phantom{\frac{1}{2}}\right.\nonumber\\
&& \phantom{W_1[}\left.\frac{1}{2}\left(\{000\}+\{001\}+\{100\}-\{101\}\right)
\right],\nonumber\\
&& J=3,\; W_2=-3W_1,\; W_3=\frac{9}{4}W_1,\; W_4=-\frac{1}{4}W_1,\;\nonumber\\
&& W_{MN}=\nonumber\\
&& W_1 \left[
\{030\}-3\{020\}+\frac{9}{4}\{010\}-\frac{1}{4}\{000\}-\right.\nonumber\\
&& \phantom{W_1[}\left.
\frac{3}{4}\left( 
\{001\}+\{100\}-\{011\}-\{110\}+\right.\right.\nonumber\\
&& \phantom{W_1[}\left.\phantom{\frac{3}{4}}\left.
\{111\}-\{101\}
\right)
\right],\label{eq:TSTmin123}
\end{eqnarray}

\subsection*{SD and DD Born Cross-sections}

Phase space and cross-section for SD:
\begin{eqnarray}
&& d\sigma^J_{{\mathrm SD}}=\frac{1}{2s}\frac{d^3\vec{k}^{\prime}}{(2\pi)^3 2k^{\prime}_0} dM^2\;\delta((p+k-k^{\prime})^2-M^2)\times\nonumber\\
&& \phantom{d\sigma}
\frac{T^{\mu_1...\mu_J}(k,q)T^{\nu_1...\nu_J}(k,q)W_{\mu_1...\nu_J}(p,q)}{(t-m_J^2)^2}=\nonumber\\
&& \phantom{d\sigma} 
\frac{dt\;d\xi}{32\pi^2 s} \frac{T^{\mu_1...\mu_J}(k,q)T^{\nu_1...\nu_J}(k,q)W_{\mu_1...\nu_J}(p,q)}{(t-m_J^2)^2},\;\label{eq:csSD1}
\\
&&\phantom{XXXXX}\nonumber\\
&& \sigma_{pJ}|_{t\to 0}=\frac{1}{2J+1}\frac{1}{2M^2}
\sum\epsilon_{\mu_1...\mu_J}\epsilon^*_{\nu_1...\nu_J}
W^{\mu_1...\nu_J}=\nonumber\\
&& \phantom{\sigma}\frac{1}{2(2J+1)M^2}\prod\limits_{i=1}^J g_{\mu_i\nu_i}W^{\mu_1...\nu_J}=\frac{2W_1(p,q)}{2(2J+1)M^2},\label{eq:sigpJ1}
\\
&&\phantom{XXXXX}\nonumber\\
&& T^{\mu_1...\mu_J}T^{\nu_1...\nu_J}W_{\mu_1...\nu_J}|_{t\to 0}=\nonumber\\
&& \phantom{F(t)}\left|F_J(t)\right|^2
\left( \frac{m^2}{m^2+|t|/4}\right)^J
\left( \frac{|t|}{|t|_{\mathrm min}}-1\right)^J W_1(p,q)=\nonumber
\end{eqnarray}
\begin{eqnarray}
&& \phantom{F(t)}\left|F_J(t)\right|^2
\left( \frac{m^2}{(m^2+|t|/4)}\right)^J\frac{1}{2^{J-1}}
\left( \frac{|t|}{|t|_{\mathrm min}}-1\right)^J\times\nonumber\\
&& \phantom{F(t)}(2J+1)M^2\sigma_{pJ}.\label{eq:LWSD1}
\end{eqnarray}

After reggeization:
\begin{eqnarray}
&&\frac{1}{t-m_J^2}\to\pi\alpha_{{\mathbb P}}^{\prime}(t)\eta_{{\mathbb P}}(t),\nonumber\\
&& J\to\alpha_{{\mathbb P}}(t),\; \eta_{{\mathbb P}}(t)={\mathrm i}+\tan\frac{\pi(\alpha_{{\mathbb P}}(t)-1)}{2},
\end{eqnarray}

the cross-section for SD looks like
\begin{eqnarray}
&& \frac{d^2\sigma^J_{{\mathrm SD}}}{d|t|\,d\xi}=\frac{1}{16\pi^2}\frac{1}{2s}\frac{1}{(t-m_J^2)^2}\times\nonumber\\
&& \phantom{\frac{d^2\sigma^J_{{\mathrm SD}}}{d|t|\,d\xi}=}
\left|F_J(t)\right|^2\left( \frac{m^2}{m^2+|t|/4}\right)^J
\frac{1}{2^{J-1}}\times\nonumber\\
&& \phantom{\frac{d^2\sigma^J_{{\mathrm SD}}}{d|t|\,d\xi}=}
\left( \frac{|t|}{|t|_{\mathrm min}}-1\right)^J (2J+1)M^2\sigma_{pJ}(M^2),\label{eq:csSD2}\\
&& \phantom{XXXXXXX}\nonumber\\
&& \frac{d^2\sigma_{{\mathrm SD}}}{d|t|\,d\xi}=\frac{(\pi\alpha_{{\mathbb P}}^{\prime}(t))^2|\eta_{{\mathbb P}}(t)|^2}{32\pi^2}\xi\times\nonumber\\
&& \phantom{\frac{d^2\sigma^J_{{\mathrm SD}}}{d|t|\,d\xi}=}
\left|F_{\alpha_{{\mathbb P}}}(t)\right|^2
\left( \frac{m^2}{(m^2+|t|/4)}\right)^{\alpha_{{\mathbb P}}(t)}
\frac{1}{2^{\alpha_{{\mathbb P}}(t)-1}}\times\nonumber\\
&& \phantom{\frac{d^2\sigma^J_{{\mathrm SD}}}{d|t|\,d\xi}=}
\left( \frac{|t|}{|t|_{\mathrm min}}-1\right)^{\alpha_{{\mathbb P}}(t)}
(2\alpha_{{\mathbb P}}(t)+1)
\sigma_{p{\mathbb P}}(M^2).\label{eq:csSD3} 
\end{eqnarray}

To obtain $F_{\alpha_{{\mathbb P}}}(t)$ we can use 
the elastic $p p $ - scattering (with J-exchange and further reggeization):
\begin{eqnarray}
 \Omega_J&=&-{\mathrm i}\frac{\left[ F_J(t)\right]^2 \left( \frac{s}{2(m^2-t/4)}\right)^J}{t-m_J^2},\label{eq:eikJ1}\\
 \Omega_{{\mathbb P}}(s,t)&=&-{\mathrm i}\pi\alpha_{{\mathbb P}}^{\prime}(t)\eta_{{\mathbb P}}(t)\left[ F_{\alpha_{{\mathbb P}}}(t)\right]^2\times\nonumber\\ 
&\phantom{=}&\left( \frac{s}{2(m^2-t/4)}\right)^{\alpha_{{\mathbb P}}(t)},
\label{eq:eikP1}
\end{eqnarray}
${\mathrm dim}\left[F_J\right]={\mathrm GeV}$.

The standard form for the eikonal now is 
\begin{eqnarray}
&& \Omega_{{\mathbb P}}(s,t)=-{\mathrm i}\pi\alpha_{{\mathbb P}}^{\prime}(t)
\eta_{{\mathbb P}}(t)g_{{\mathbb P}}(t)^2
\left( \frac{s}{s_0}\right)^{\alpha_{{\mathbb P}}(t)}\label{eq:eikusual}\\
&& F_{\alpha_{{\mathbb P}}}(t)
\left( \frac{s_0}{2(m^2-t/4)}\right)^{\alpha_{{\mathbb P}}(t)/2}=
g_{{\mathbb P}}(t),\label{eq:gFrelation}
\end{eqnarray}
where $g_{{\mathbb P}}(t)$ is the proton-proton-Pomeron vertex coupling.

Phase space and differential cross-section for DD are:
\begin{eqnarray}
&& d\sigma^J_{{\mathrm DD}}=\frac{1}{2s}\frac{d^4\vec{k}^{\prime}}{(2\pi)^4} 
dM_1^2dM_2^2\times\nonumber\\
&&\phantom{d\sigma^J_{{\mathrm DD}}=}\delta((p+q)^2-M_2^2)\,\delta((k-q)^2-M_1^2)\times\nonumber\\
&& \phantom{d\sigma^J_{{\mathrm DD}}=}\frac{W^{\mu_1...\nu_J}(k,-q)W_{\mu_1...\nu_J}(p,q)}{(t-m_J^2)^2}=\nonumber\\
&& \phantom{d\sigma^J_{{\mathrm DD}}=}\frac{dt\,d\xi_1\,d\xi_2}{64\pi^3} 
\frac{W^{\mu_1...\nu_J}(k,-q)W_{\mu_1...\nu_J}(p,q)}{(t-m_J^2)^2},
\;\label{eq:csDD1}
\end{eqnarray}
\begin{eqnarray}
&& \left. W^{\mu_1...\nu_J}(k,-q)W_{\mu_1...\nu_J}(p,q)
\right|_{|t|\ll m^2\ll M_i^2}\simeq
\nonumber\\
&& \phantom{d\sigma^J_{{\mathrm DD}}=}4\left( \frac{|t|}{|t|_{\mathrm min}}\right)^{2J}
\left[ \frac{1+(1-|t|_{\mathrm min}/|t|)^{2J}}{2}\right]\times\nonumber\\
&& \phantom{....}\nonumber\\
&& \phantom{d\sigma^J_{{\mathrm DD}}=}W_1(k,-q)W_1(p,q),\label{eq:WWDD}
\end{eqnarray}
where the factor in the square brackets is equal to unity for \linebreak$|t|\gg |t|_{\mathrm min}$ 
and to $1/2$ at $|t|=|t|_{\mathrm min}$ (the overall factor in front of $W_1(k,-q)W_1(p,q)$ is equal to $2$ in this case), $|t|_{\mathrm min}\simeq \xi_1\xi_2 s$.

Now we can write down the cross-section for DD:
\begin{eqnarray}
&& \frac{d^3\sigma^J_{DD}}{d|t|\,d\xi_1\,d\xi_2}=
\frac{s^2\xi_1\xi_2}{64\pi^3}\frac{1}{(t-m_J^2)^2}\times\nonumber\\
&& \phantom{\frac{d^3\sigma^J_{DD}}{d|t|\,d\xi_1\,d\xi_2}}
4\left( \frac{|t|}{|t|_{\mathrm min}}\right)^{2J}
\left[ \frac{1+(1-|t|_{\mathrm min}/|t|)^{2J}}{2}\right]\times\nonumber\\
&& \phantom{\frac{d^3\sigma^J_{DD}}{d|t|\,d\xi_1\,d\xi_2}}
(2J+1)^2\sigma_{pJ}(M_1^2)\sigma_{pJ}(M_2^2),\label{eq:csDDpJ}\\
&& \phantom{XXXXXX}\nonumber\\
&& \frac{d^3\sigma_{DD}}{d|t|\,d\xi_1\,d\xi_2}=
\frac{(\pi\alpha_{{\mathbb P}}^{\prime}(t))^2|\eta_{{\mathbb P}}(t)|^2}{64\pi^3}
s^2\xi_1\xi_2\times\nonumber\\
&& \phantom{\frac{d^3\sigma^J_{DD}}{d|t|\,d\xi_1\,d\xi_2}}
4\left( \frac{|t|}{|t|_{\mathrm min}}\right)^{2{\alpha_{{\mathbb P}}(t)}}
\left[ \frac{1+(1-|t|_{\mathrm min}/|t|)^{2{\alpha_{{\mathbb P}}(t)}}}{2}\right]
\times\nonumber\\
&&\phantom{\frac{d^3\sigma^J_{DD}}{d|t|\,d\xi_1\,d\xi_2}}
(2{\alpha_{{\mathbb P}}(t)}+1)^2\sigma_{p{\mathbb P}}(M_1^2)\sigma_{p{\mathbb P}}(M_2^2).\label{eq:csDDpP} 
\end{eqnarray}

\subsection*{Rescattering corrections and ways to extract the proton-Pomeron cross-section from the data}

Here we consider several cases in which we could extract
proton-Pomeron cross-sections from the data on SD and DD.

Let us introduce new notations:
\begin{eqnarray}
&& {\cal F}_{SD}(t,\xi)=\nonumber\\
&& \phantom{{\cal F}_{SD}}
\frac{\alpha_{{\mathbb P}}^{\prime}(t)\eta_{{\mathbb P}}(t)}{4}
\sqrt{\xi}
F_{\alpha_{{\mathbb P}}}(t)
\left( \frac{m^2}{2(m^2+|t|/4)}\right)^{\alpha_{{\mathbb P}}(t)/2}
\times\nonumber\\
&& \phantom{{\cal F}_{SD}}\left( \frac{|t|}{|t|_{\mathrm min}}-1\right)^{\alpha_{{\mathbb P}}(t)/2}
\sqrt{2\alpha_{{\mathbb P}}(t)+1}\label{eq:fluxSDt}
\end{eqnarray}
and
\begin{eqnarray}
&& {\cal F}_{DD}(t,\xi_1,\xi_2)=\nonumber\\
&& \phantom{{\cal F}_{DD}}
\frac{\alpha_{{\mathbb P}}^{\prime}(t)\eta_{{\mathbb P}}(t)}{4\sqrt{\pi}}
s\sqrt{\xi_1\xi_2}
\left( \frac{|t|}{|t|_{\mathrm min}}\right)^{{\alpha_{{\mathbb P}}(t)}}\times\nonumber\\
&&\phantom{{\cal F}_{DD}} 
\left[ \frac{1+(1-|t|_{\mathrm min}/|t|)^{2{\alpha_{{\mathbb P}}(t)}}}{2}\right]^{1/2}
(2\alpha_{{\mathbb P}}(t)+1).
\label{eq:fluxDDt}
\end{eqnarray}
Then we have from (\ref{eq:csSD3}) and (\ref{eq:csDDpP}):
\begin{eqnarray}
\hspace*{-0.6cm} \frac{d^2\sigma_{SD}}{d|t|\,d\xi}&=&\left| {\cal F}_{SD}(t,\xi)\right|^2 
\sigma_{p{\mathbb P}}(M^2;t),\label{eq:csSDshort}\\
\hspace*{-0.6cm} \frac{d^3\sigma_{DD}}{d|t|\,d\xi_1\,d\xi_2}&=&\left| {\cal F}_{DD}(t,\xi_1,\xi_2)\right|^2 
\sigma_{p{\mathbb P}}(M_1^2;t)\sigma_{p{\mathbb P}}(M_2^2;t).\label{eq:csDDshort}
\end{eqnarray}
Here we explicitly indicate generally present $t$-dependence of the proton-Pomeron cross-section. Further on it will be seen to be important for unitarization.

Let us assume first that the proton-Pomeron total cross-section can be
represented in a factorized form:
\begin{equation}
\sigma_{p{\mathbb P}}(M^2;t)=h(M^2)f(t).\label{eq:assumesigma}
\end{equation}
Now we can take into account rescattering corrections which are depicted
as V-blobes in Fig.~\ref{fig1:processes}. To this end it is convenient
to absorb all the $t$-dependent functions into a single one:
\begin{eqnarray}
 {\cal H}_{SD}(t,\xi)&=&{\cal F}_{SD}(t,\xi)\sqrt{f(t)},\label{eq:HSD}\\
 {\cal H}_{DD}(t,\xi_1,\xi_2)&=&{\cal F}_{DD}(t,\xi_1,\xi_2)f(t)\label{eq:HDD}.
\end{eqnarray}
Then the unitarization procedure will transform these functions to
\begin{equation}
\label{eq:HUni}
{\cal H}_a^U(t;\dots)=\int\limits_0^{\infty}b\,db\,J_0(b\sqrt{|t|})\,
{\mathrm e}^{-\tilde{\Omega}_{{\mathbb P}}(s,b)}\,\tilde{{\cal H}}_a(b;\dots),
\end{equation}
where $a=SD,\,DD$, and
\begin{eqnarray}
\tilde{{\cal H}}_a(b;\dots)&=&\int\limits_0^{\infty}
\tau\,d\tau\,J_0(b\,\tau){\cal H}_a(-\tau^2;\dots), \label{eq:Hab}\\
\tilde{\Omega}_{{\mathbb P}}(s,b)&=&\frac{1}{16\,\pi\,s}\int\limits_0^{\infty}
\tau\,d\tau\,J_0(b\,\tau)\,\Omega_{{\mathbb P}}(s,-\tau^2),\label{eq:Omegab}
\end{eqnarray}
$\tau=\sqrt{|t|}$, $\Omega_{{\mathbb P}}(s,t)$ is 
taken from~(\ref{eq:eikusual}).

\begin{figure}[t!]  
 \includegraphics[width=0.49\textwidth]{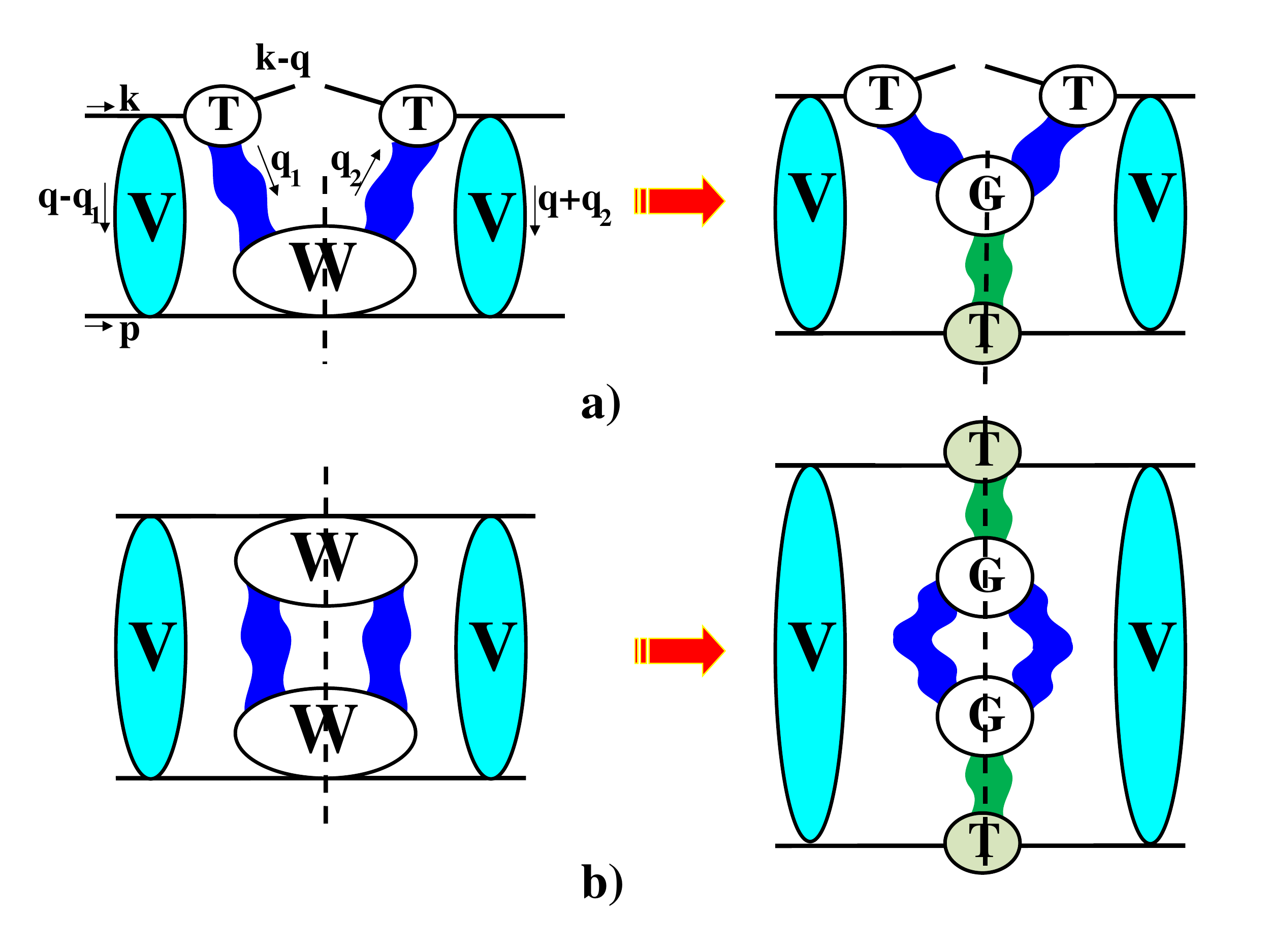} 
  \caption{\label{fig2:unitar} Exact scheme of unitarization in the case of SD (a) and DD (b) cross-sections. Right pictures represent the case with 3-Pomeron vertices G.}
\end{figure}
Strictly speaking, the exact expression with absorption looks as (see Fig.\ref{fig2:unitar})
\begin{eqnarray}
&&\hspace*{-0.2cm}
{\cal H}_a^{(2)\,U}(-\vec{q}^2;\dots)=\frac{1}{(4\pi)^2}
\int\limits_0^{\infty}\int\limits_0^{\infty}
d\vec{q}_1^2\,d\vec{q}_2^2\,
\int\limits_0^{2\pi}\int\limits_0^{2\pi}
d\phi_1\, d\phi_2 \, \nonumber\\
&&\hspace*{-0.2cm} V\left(s,(\vec{q}-\vec{q}_1)^2\right)
{\cal H}_a^{(2), nf}\left(\vec{q}_1^2,\vec{q}_2^2,\vec{q}_1\vec{q}_2;\dots\right)
V\left(s,(\vec{q}+\vec{q}_2)^2\right), \nonumber\\
&&\hspace*{-0.2cm} V(s,\vec{q}^2)=\int\limits_0^{\infty}b\,db\,J_0(b|\vec{q}|)
{\mathrm e}^{-\tilde{\Omega}_{{\mathbb P}}(s,b)},\label{eq:H2Uni}
\end{eqnarray}
where ${\cal H}_a^{(2), nf}$ is 
the contraction of two tensors $T$ with the 
nonforward hadronic tensor (SD, Fig.\ref{fig2:unitar}(a)) or
two nonforward hadronic tensors (DD, Fig.\ref{fig2:unitar}(b)). Only 
if ${\cal H}_a^{(2), nf}$ can be represented in a factorized form
\begin{equation}
\label{eq:Hfactorized}
{\cal H}_a^{(2), nf}(t_1,t_2;\dots)={\cal H}_a(t_1;\dots){\cal H}^*_a(t_2;\dots),
\end{equation}
we will get the expression~(\ref{eq:HUni}) and
${\cal H}_a^{(2)\,U}=\left|{\cal H}_a^{U}\right|^2$. In some cases
it is rather good approximation that simplifies much the calculations.

Now final expressions for cross-sections look as
\begin{eqnarray}
\hspace*{-0.6cm} \frac{d^2\sigma^U_{SD}}{d|t|\,d\xi}&=& {\cal H}_{SD}^{(2)\,U}(t;\xi) h(M^2)\equiv\nonumber\\
\hspace*{-0.6cm} &\phantom{=}&
\frac{{\cal H}_{SD}^{(2)\,U}(t;\xi)}{f(t)} \sigma_{p{{\mathbb P}}}(M^2;t),\label{eq:csSDUh}\\
\hspace*{-0.6cm} \frac{d^3\sigma^U_{DD}}{d|t|\,d\xi_1\,d\xi_2}&=& {\cal H}_{DD}^{(2)\,U}(t;\xi_1,\xi_2)
h(M_1^2)h(M_2^2)\equiv\nonumber\\
\hspace*{-0.6cm} &\phantom{=}&
\frac{{\cal H}_{DD}^{(2)\,U}(t;\xi_1,\xi_2)}{\left[f(t)\right]^2} 
\sigma_{p{{\mathbb P}}}(M_1^2;t)\sigma_{p{{\mathbb P}}}(M_2^2;t).
\label{eq:csDDUh}
\end{eqnarray}
If we take use of some concrete model expressions for $h(M^2)$ and $f(t)$, we can extract the proton-Pomeron cross-section from the data on SD and DD with
the help of~(\ref{eq:csSDUh}) and~(\ref{eq:csDDUh}). If we know only
integrated cross-sections we can use expressions:
\begin{eqnarray}
\hspace*{-0.4cm} 
\sigma_{p{{\mathbb P}}}(M^2;t)&=&
\frac{
\frac{d\sigma^U_{SD}}{d\xi}
}{
\int\limits_{t_1}^{t_2}{\cal H}_{SD}^{(2)\,U}(t;\xi)
}f(t),
\label{eq:extractSD2}\\
\hspace*{-0.4cm} 
\sigma_{p{{\mathbb P}}}(M_1^2;t)\sigma_{p{{\mathbb P}}}(M_2^2;t)&=&
 \frac{
\frac{d^2\sigma^U_{DD}}{d\xi_1\,d\xi_2}
}{
\int\limits_{t_1}^{t_2}{\cal H}_{DD}^{(2)\,U}(t;\xi_1,\xi_2)
}\left[f(t)\right]^2.
\label{eq:extractDD2}
\end{eqnarray}
To extract the proton-Pomeron cross-section from $\sigma_{SD,\,DD}^{tot}$
we need exact expressions for $f(t)$ and $h(M^2)$, or the complete
formula for $\sigma_{p{{\mathbb P}}}(M^2;t)$ without the 
assumption~(\ref{eq:assumesigma}). 

Now we summarize our assumptions which
allow us to say something about the proton-Pomeron cross-section:
\begin{enumerate}
\item Finiteness of the hadronic tensor $W_{\mu_1\dots\nu_J}(p,q)$ for 
\linebreak
$t=q^2\to\sim\! 0$ which gives additional relations between 
structure functions and leaves us with a single function. This function
is directly related to the proton-Pomeron (totally transverse) 
cross-section through~(\ref{eq:sigpJ1}) after reg\-geiza\-tion.
\item The unitarization procedure is reduced to
rescattering corrections in the initial state. This may not be
exactly the case, but we hope that in the appropriate
kinematic region we can use this approximation.
\item Furthermore, we have to assume also the 
concrete pa\-ra\-me\-tri\-za\-tion for $\sigma_{p{{\mathbb P}}}(M^2;t)$ if we want to extract it
from SD (DD) differential (or, in the more complicated case, integrated)
cross-sections. Only if we take $f(t)\equiv 1$ and in the absence of
rescattering corrections (the very strong assumption, since
we know from the elastic scattering that for 
high energies this is not the case), we can
speak about a model-free extraction of $\sigma_{p{{\mathbb P}}}(M^2)$. 
\item Other complications can arise if we try to take into
account a contribution from secondary reggeons which can drastically spoil
the above picture. As we suppose, it can be solved by
the choice of an appropriate kinematic domain (see below).
\end{enumerate}

Finally, when we extract somehow the proton-Pomeron cross-section from
the data at different energies, we have to make 
sure that it does not depend on the overall energy ($s$). If we detect
such a dependence, it means that some of our assumptions are
wrong and should be modified.

\subsection*{Experimental data and extracted cross-section for different cases}

To start we use the following possibilities:
\begin{enumerate}
\item[I] no rescattering corrections, ${\cal H}_a^{(2)\,U}=|{\cal H}_a|^2$. In
this case we can directly extract
$\sigma_{p{{\mathbb P}}}(M^2,t)$ without any assumptions
about its behaviour;
\item[II] rescattering is taken into account, $f(t)\equiv 1$;
\item[III] rescattering is taken into account, $f(t)=t^{-\alpha_{{\mathbb P}}(0)/2}$; 
\item[IV] rescattering is taken into account,  $f(t)=t^{-\alpha_{{\mathbb P}}(t)}$
\end{enumerate}
and apply them against the data from~\cite{data1}-\cite{data6}.

To be sure that the contribution from secondary reggeons is negligible  we consider the 
following (``Pomeron dominated'') kinematic region:
\begin{itemize}
\item $t$ 
should be rather small to express the hadronic tensor in terms of a single function:
\begin{equation}
0.01\,{\mathrm GeV}^2<|t|<0.1\,{\mathrm GeV}^2\,\label{eq:kinmod0}\\
\end{equation}
\item to exclude contributions from secondary reggeons (which contribute less than 15\% at $\sqrt{s}>\sqrt{s_{\mbox{{\tiny ISR}}}}=62$~GeV in the elastic scattering) we can use the
``rapidity gap'' condition
\begin{eqnarray}
&& \ln\frac{s}{M^2}>\ln\frac{s_{\mbox{{\tiny ISR}}}}{2m^2}\rightarrow\,\xi<0.00052\mbox{ (SD)},\nonumber\\
&& M<0.023\sqrt{s}\sim 160\,{\mathrm GeV}\mbox{ at 7~TeV LHC},
\label{eq:kinmod1}\\
&& \ln\frac{s s_0}{M_1^2M_2^2}>\ln\frac{s_{\mbox{{\tiny ISR}}}}{2m^2}\rightarrow
\sqrt{\xi_1\xi_2}<\sqrt{\frac{2m^2s_0}{s s_{\mbox{{\tiny ISR}}}}}\simeq \frac{0.023}{\sqrt{s}}\mbox{ (DD)}\nonumber\\
&& \sqrt{M_1M_2}<\sqrt{0.023\sqrt{s}}\sim 12.6\,{\mathrm GeV}\mbox{ at 
7~TeV LHC}\label{eq:kinmod2}
\end{eqnarray}
\item if we use 3-Pomeron vertex then (in the case of conserved tensor currents)
\begin{eqnarray}
&& \frac{M_i}{(|t|m^2)^{1/4}}>\sqrt{\frac{s_{\mbox{{\tiny ISR}}}}{2}}\sim
43.\label{eq:kinmod3}
\end{eqnarray}
\end{itemize}

For DD we have much
less possibilities to choose appropriate kinematics
for cases IIa,b: even for very large $\sqrt{s}\sim 10$~TeV 
and rather low $|t|\sim 0.01$~GeV$^2$, i. e. 
it will be difficult to extract $\sigma_{p{{\mathbb P}}}$ from 
cross-sections, integrated, for example, in $\xi_{1,2}$ ($M_{1,2}$). 


Below we consider a possibility to extract proton-Po\-me\-ron cross-section
from the existing data on SD. Before starting this task let us note
several important facts:
\begin{itemize}
\item The first one concerns the conservation of 
tensor currents, the basic ingredients 
of our calculations. In the case of conserved
currents we can obtain classical Regge formulae
in the most natural way. We immediately
obtain Legendre polynomials in the elastic scattering,
with all the scales exactly defined in the argument
of these polynomials. The specific feature of this approach
is the special $t$-dependence of differential 
cross-sections in the case of unequal masses,\linebreak 
SD~(\ref{eq:csSD3}) 
or DD~(\ref{eq:csDDpP}). This behaviour
at small t can spoil a description of the
data and lead to the strong $t$-dependence 
of the proton-Po\-me\-ron cross-section
as is seen in figures of this section.
\item The second one is the need in a correct procedure of 
the unitarization. This becomes 
clear if we try to extract the proton-Pomeron cross-section
at different values of the ``alien'' energy $\sqrt{s}$. This is the case I in our 
tests, and it is depicted in 
Figs.~\ref{fig:sigpPt}-\ref{fig:sigpPM3}(a). It is seen that
if we extract the proton-Po\-me\-ron cross-section without 
unitarization it turns out to be $s$-dependent  which 
is completely unacceptable.

\begin{figure}[hbt!]  
 \includegraphics[width=0.22\textwidth]{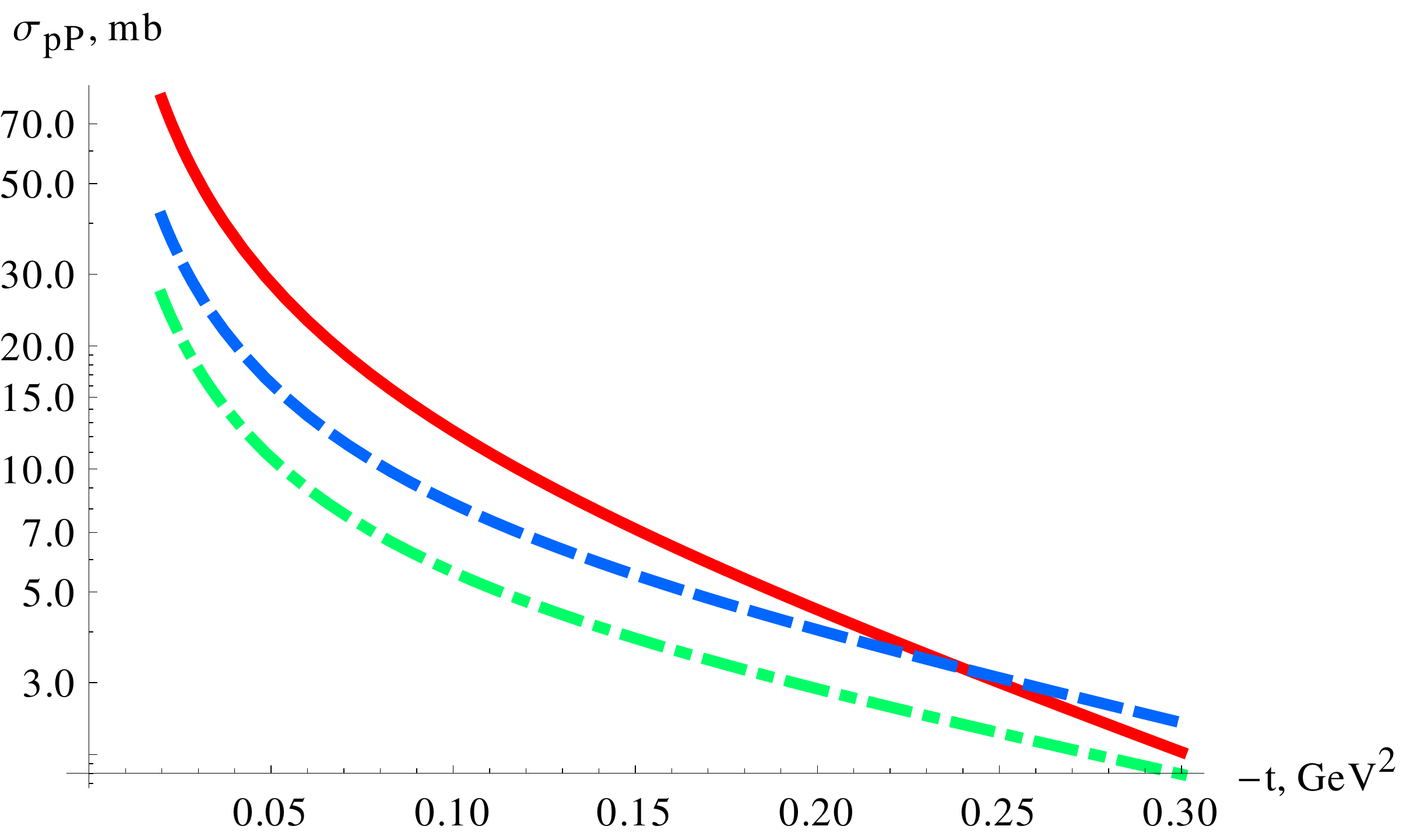} 
 \includegraphics[width=0.22\textwidth]{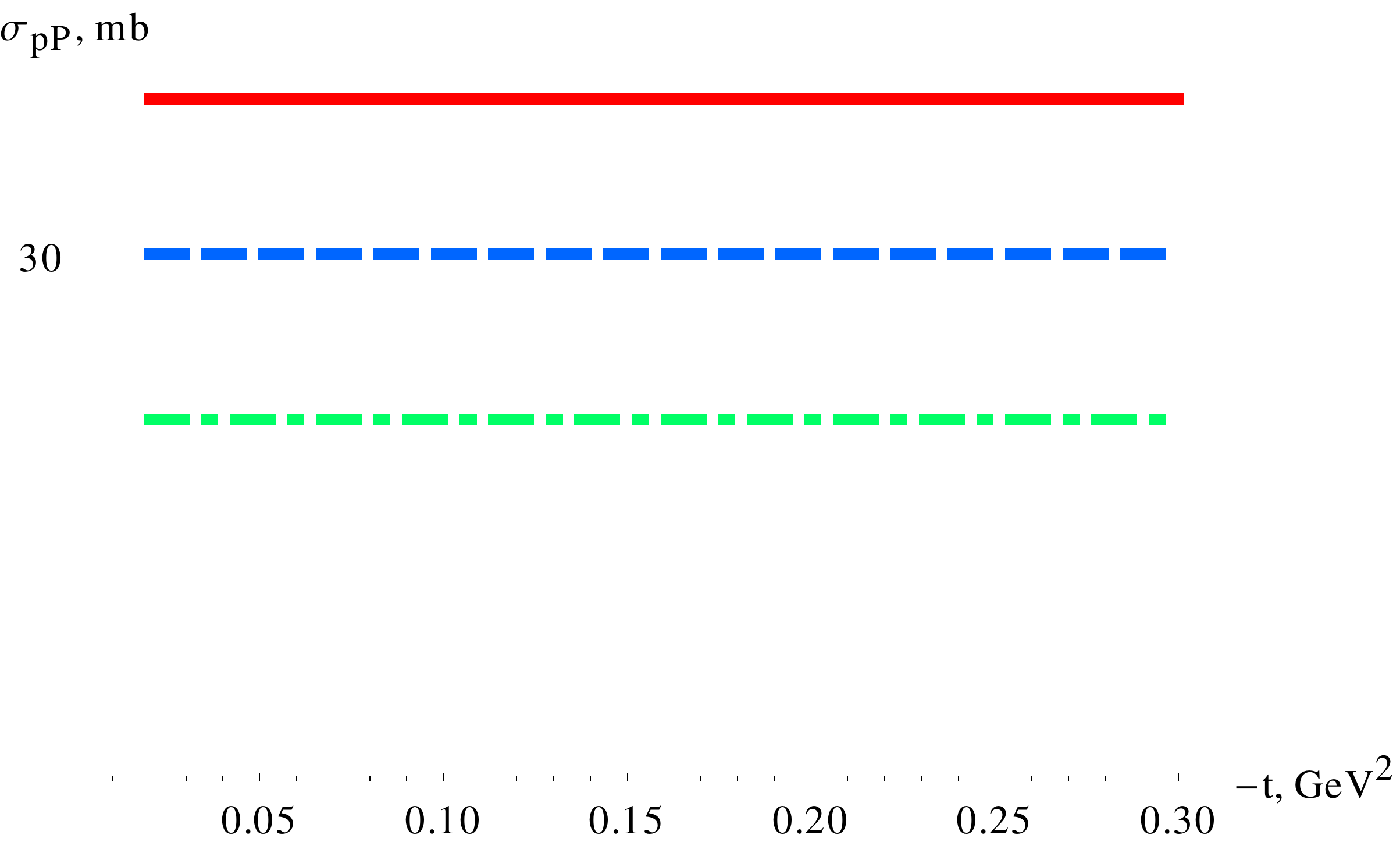}\\
 \hspace*{0.11\textwidth}a)\hspace*{0.20\textwidth}b)\\
 \includegraphics[width=0.22\textwidth]{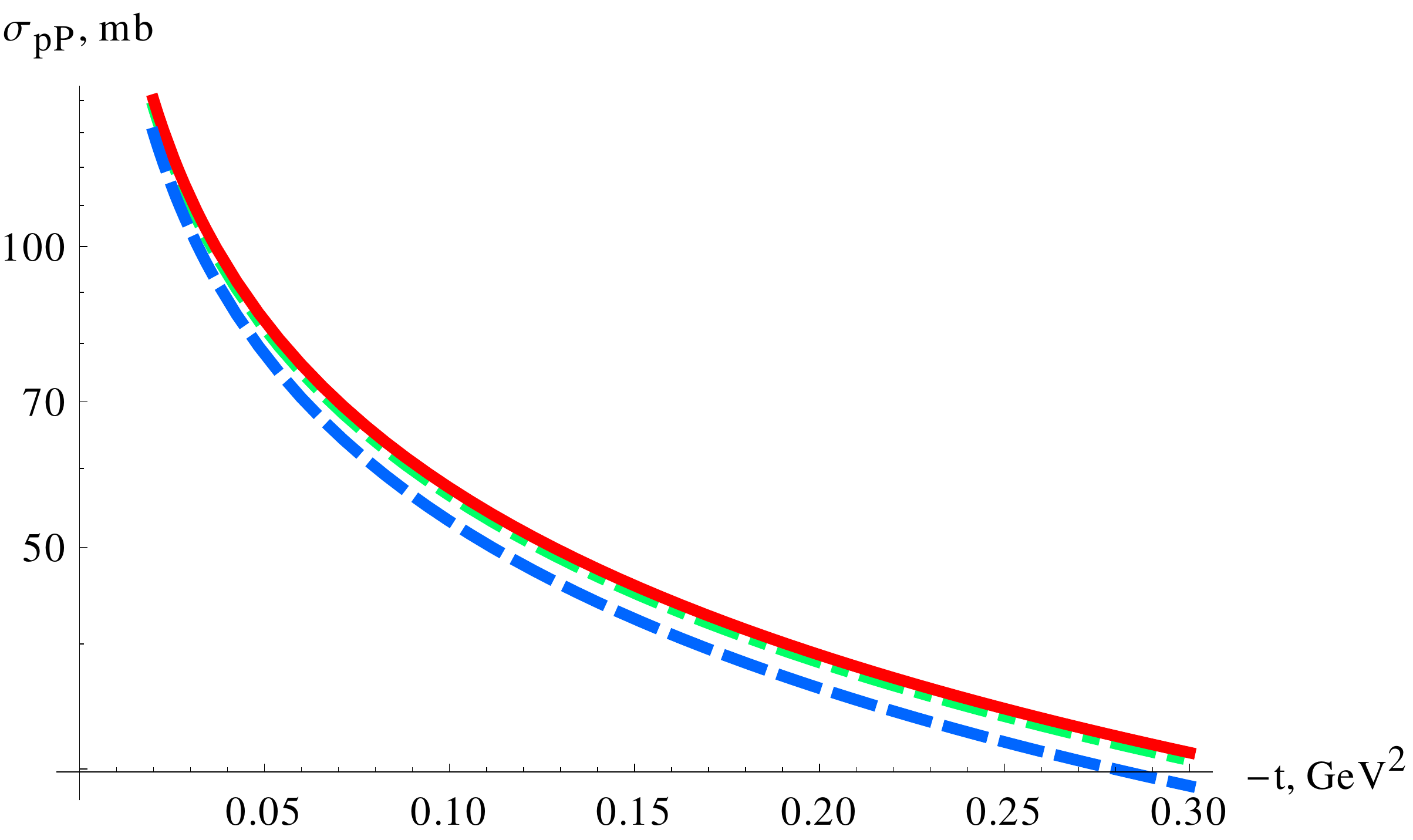}
 \includegraphics[width=0.22\textwidth]{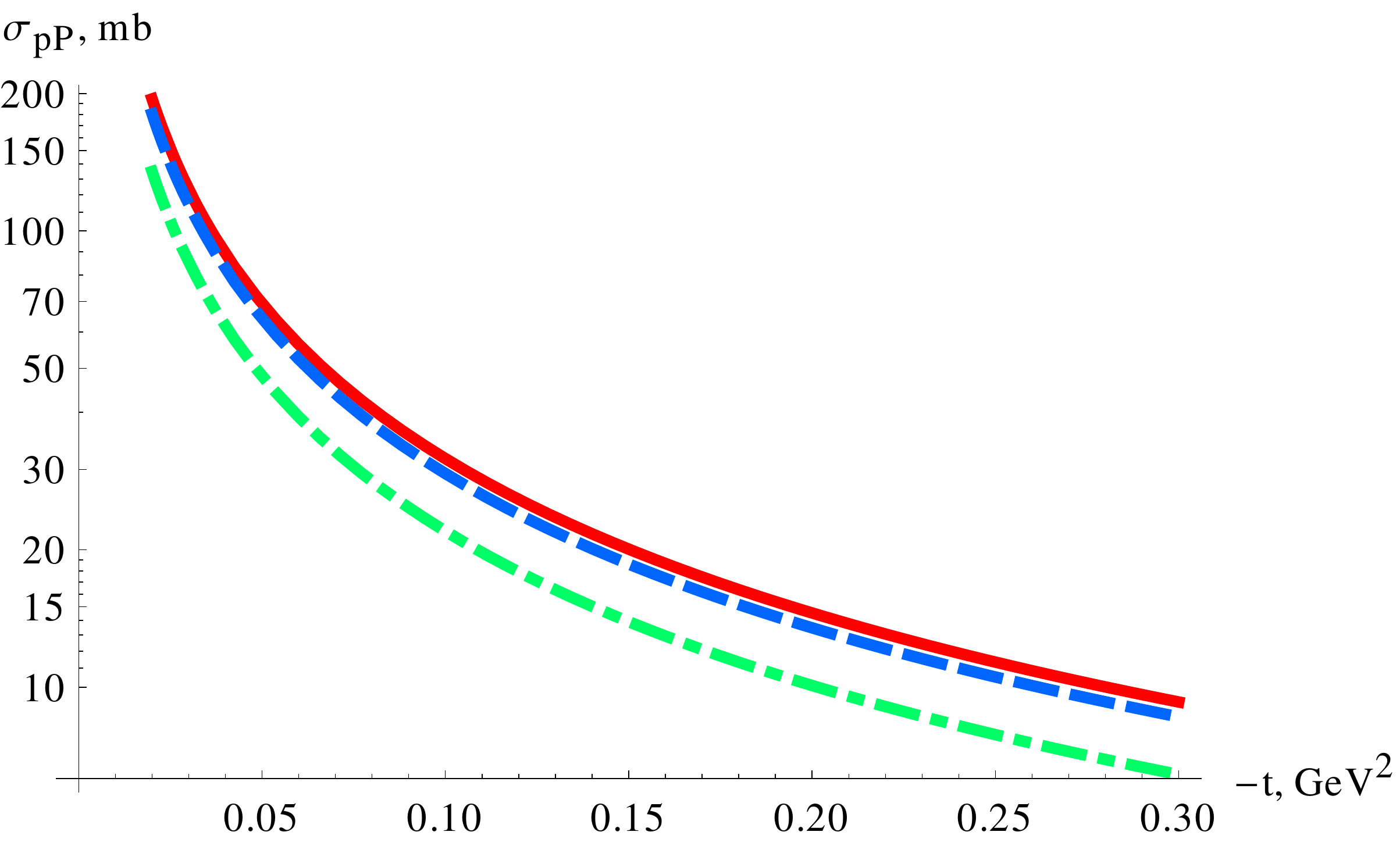}\\
 \hspace*{0.11\textwidth}c)\hspace*{0.20\textwidth}d)
  \caption{\label{fig:sigpPt} $t $-dependence of the extracted total Pomeron-proton cross-sections at $M=20$~GeV for different 
  cases: a) I; b) II; c) III; d) IV. Related semi-quantitative description of the data is depicted in 
  Figs.~\ref{fig:dsigSDCDFII}-\ref{fig:dsigSDTOTEMt}. Solid curves represent the data at 
  $\sqrt{s}=546$~GeV, dashed curves are related to $\sqrt{s}=1800$~GeV, and dash-dotted are related to $\sqrt{s}=7$~TeV.}
\end{figure}

\begin{figure}[hbt!]  
 \includegraphics[width=0.22\textwidth]{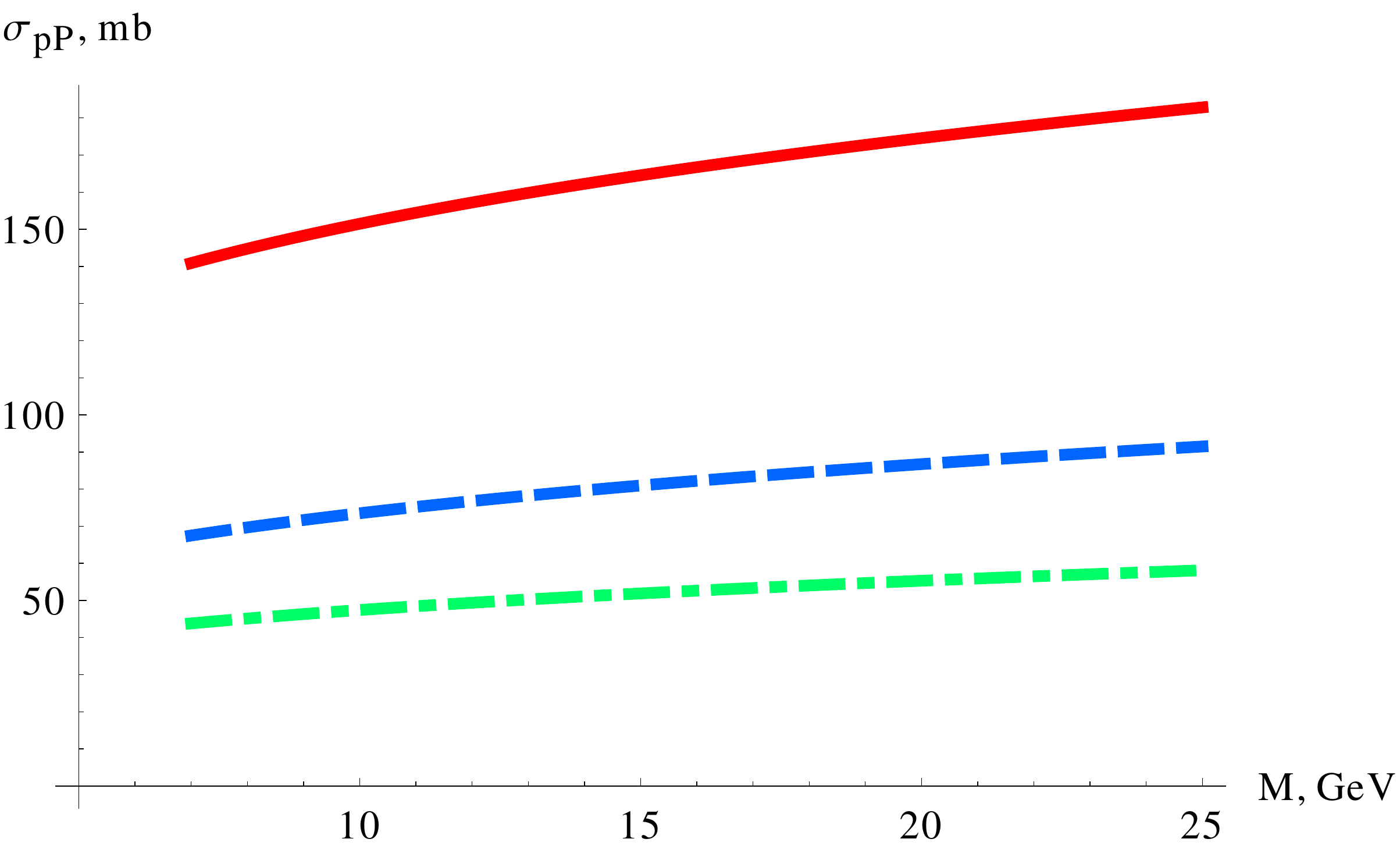} 
 \includegraphics[width=0.22\textwidth]{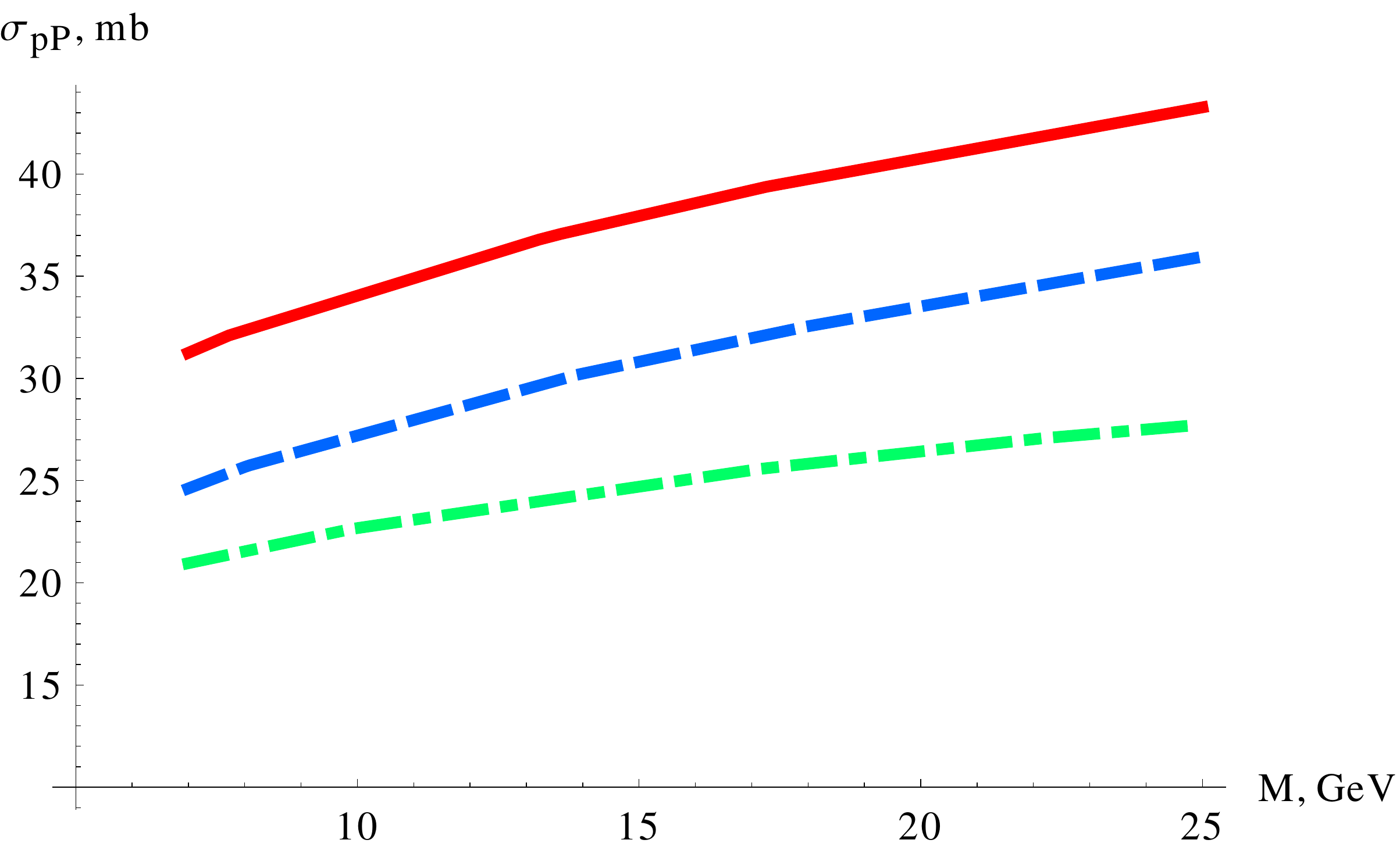}\\
 \hspace*{0.11\textwidth}a)\hspace*{0.20\textwidth}b)\\
 \includegraphics[width=0.22\textwidth]{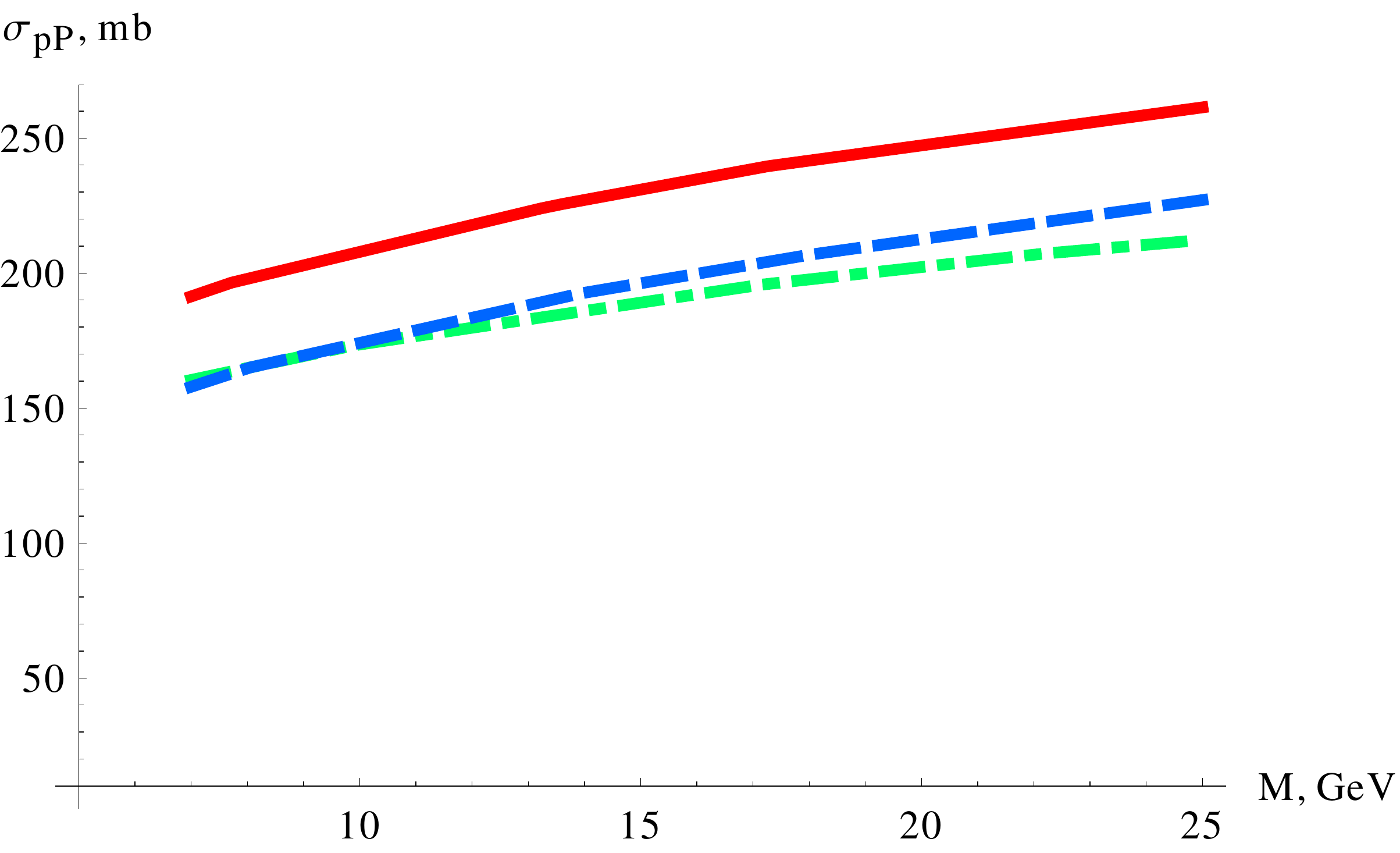}
 \includegraphics[width=0.22\textwidth]{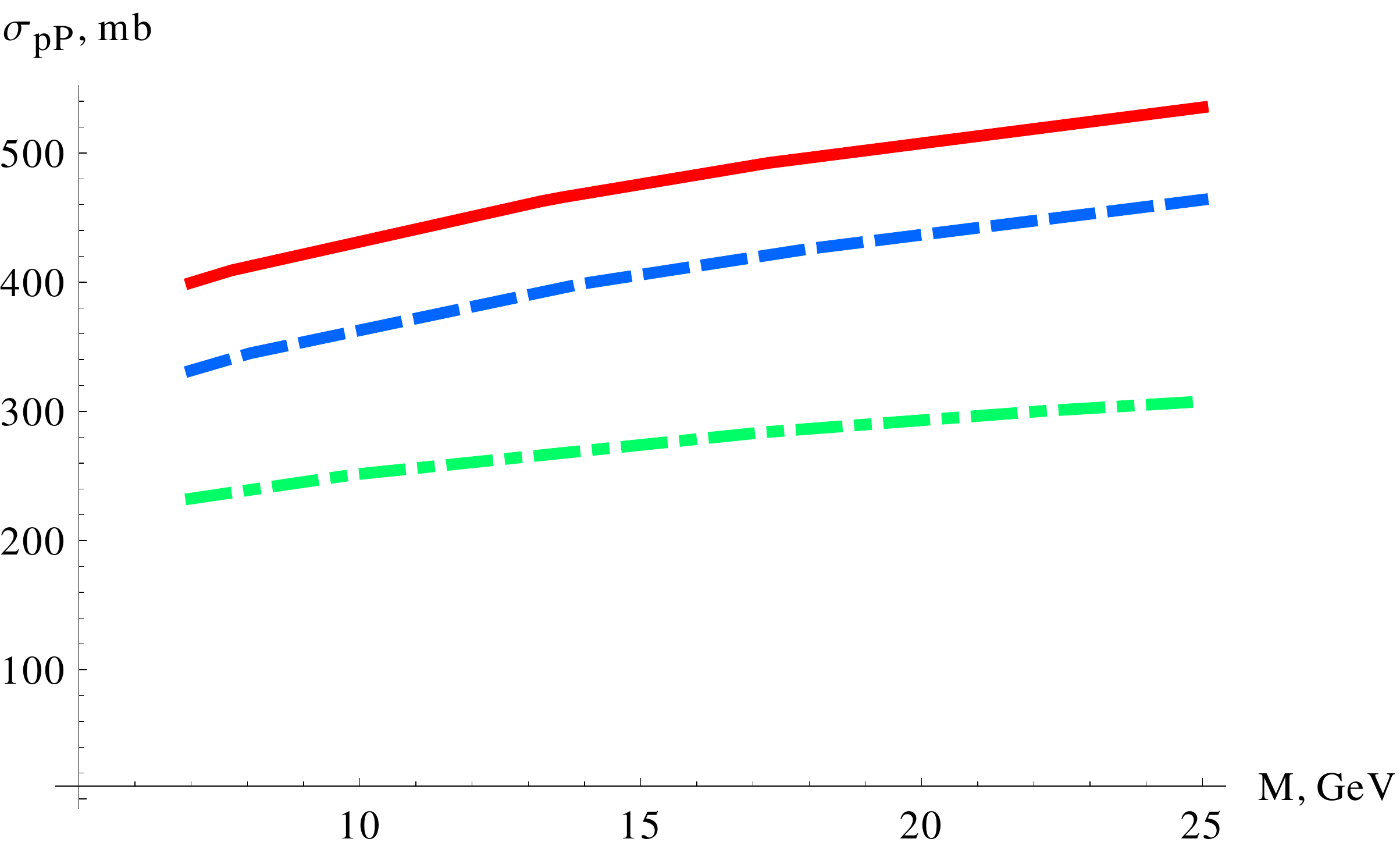}\\
 \hspace*{0.11\textwidth}c)\hspace*{0.20\textwidth}d)
  \caption{\label{fig:sigpPM1} M-dependence of the extracted total Pomeron-proton cross-sections for different cases: a) I; b) II; c) III; d) IV at $t=-0.01\,\mbox{GeV}^2$. Related semi-quantitative description of the data is depicted in Figs.~\ref{fig:dsigSDCDFII}-\ref{fig:dsigSDTOTEMt}. Solid curves represent the data at $\sqrt{s}=546$~GeV, dashed curves are related to $\sqrt{s}=1800$~GeV, and dash-dotted are related to $\sqrt{s}=7$~TeV.}
\end{figure}

\begin{figure}[hbt!]  
 \includegraphics[width=0.22\textwidth]{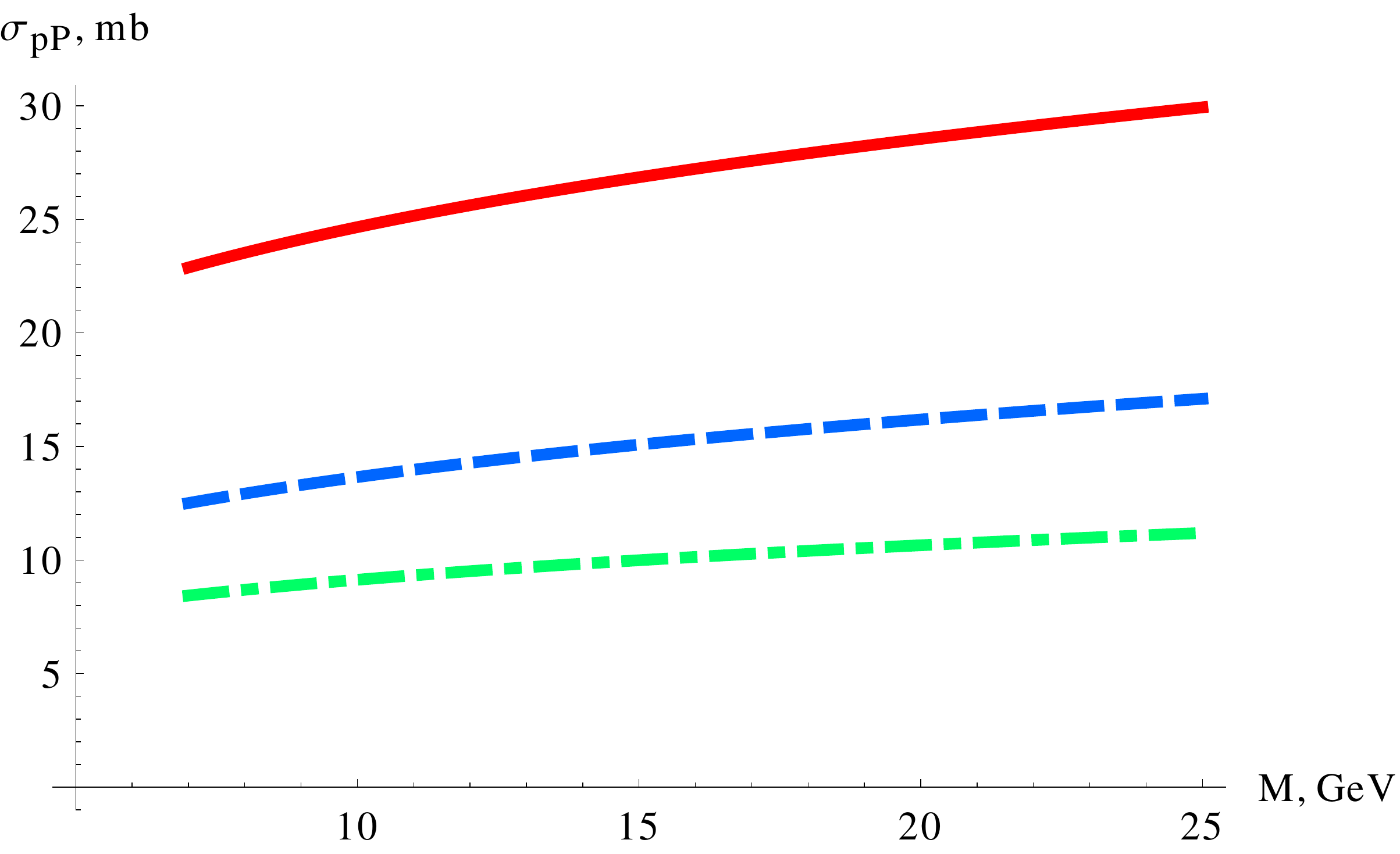} 
 \includegraphics[width=0.22\textwidth]{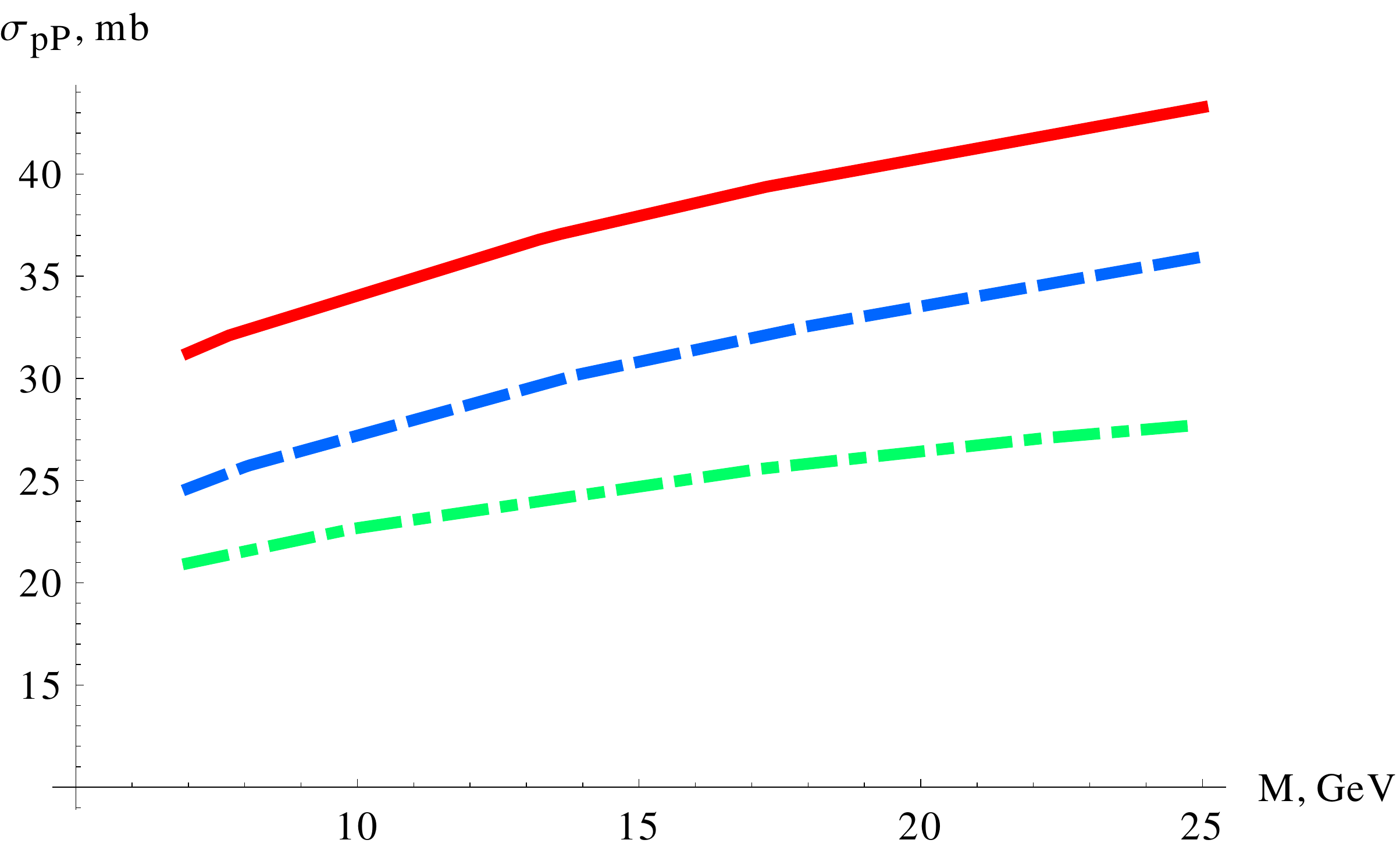}\\
 \hspace*{0.11\textwidth}a)\hspace*{0.20\textwidth}b)\\
 \includegraphics[width=0.22\textwidth]{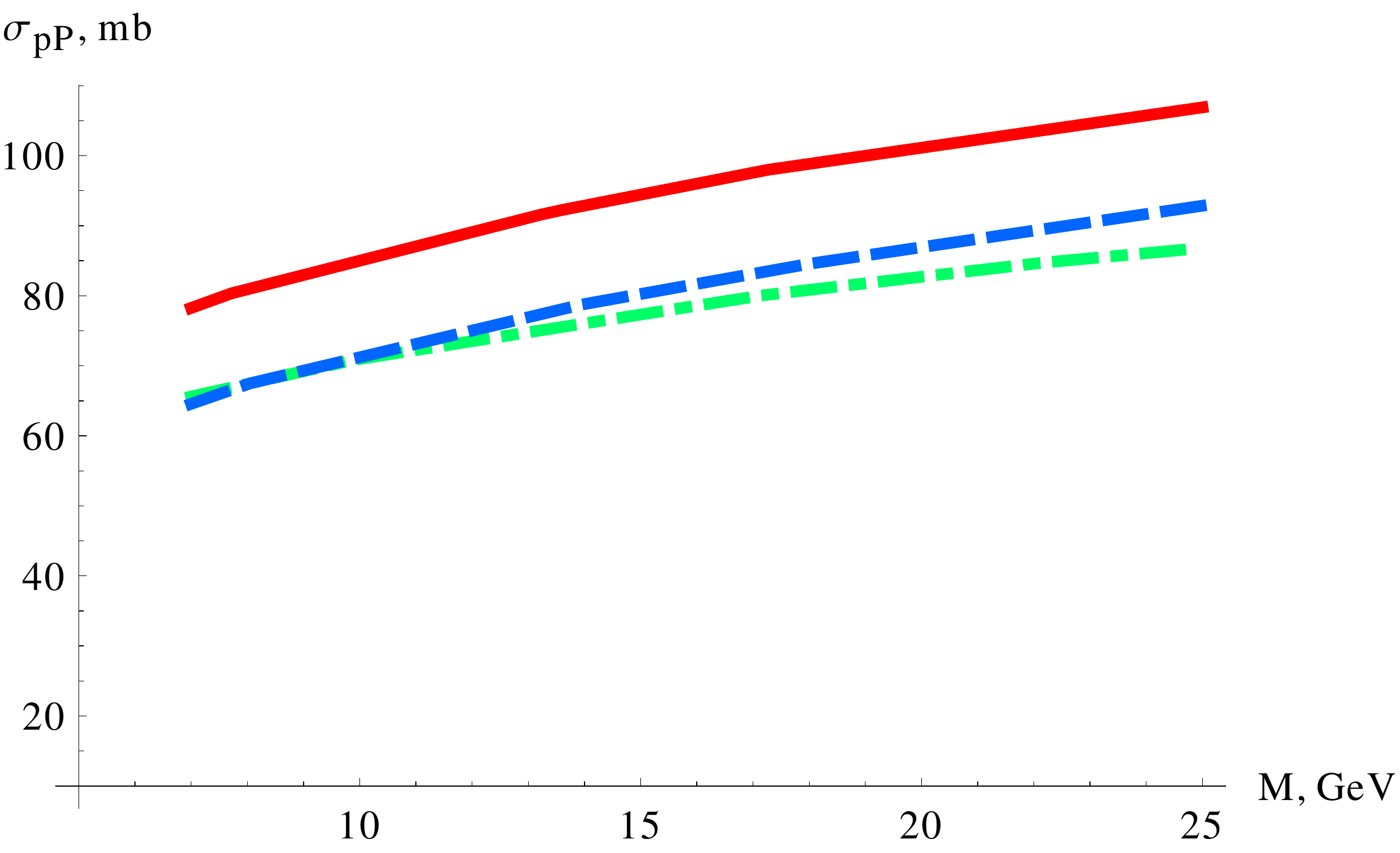}
 \includegraphics[width=0.22\textwidth]{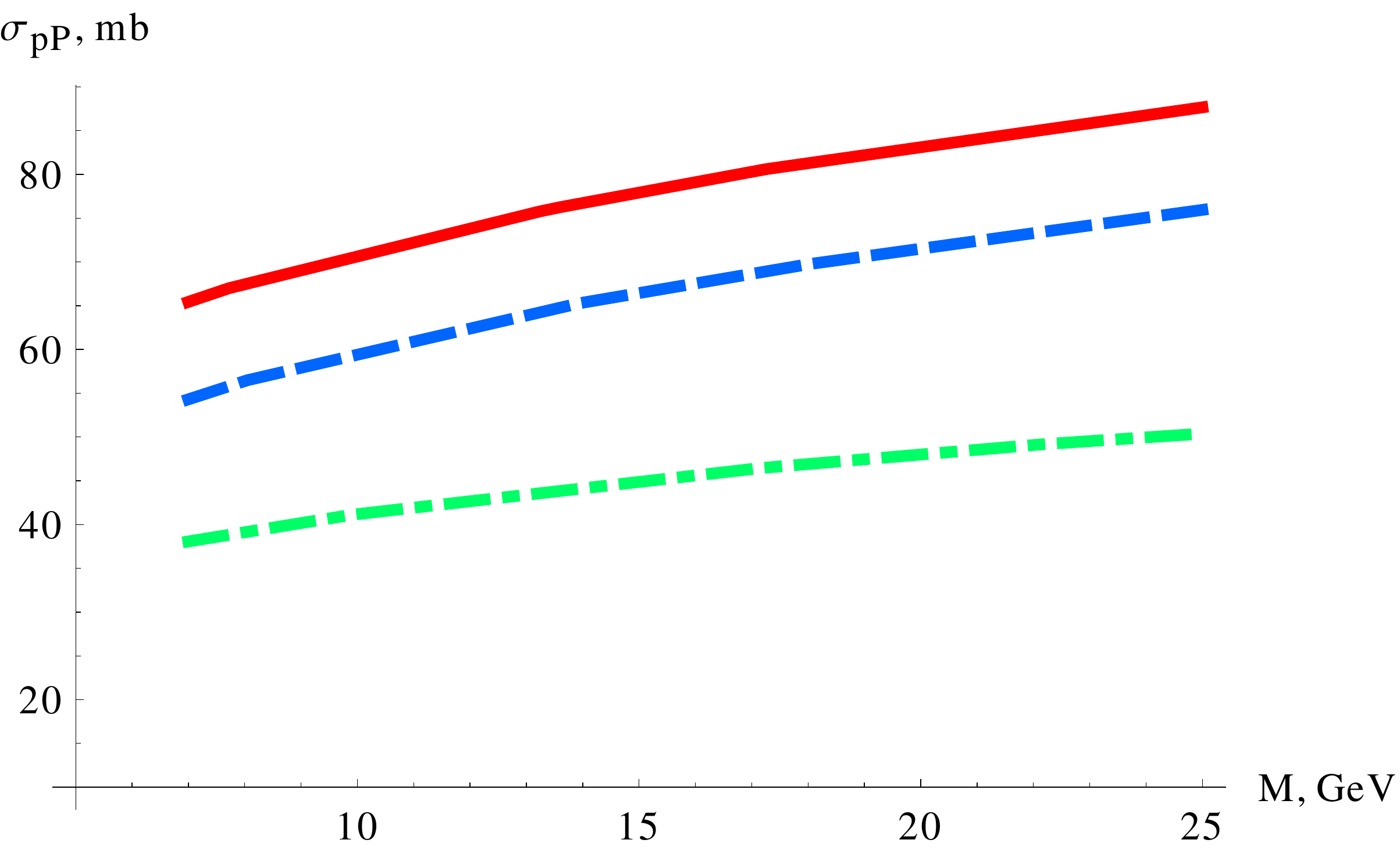}\\
 \hspace*{0.11\textwidth}c)\hspace*{0.20\textwidth}d)
  \caption{\label{fig:sigpPM2} M-dependence of the extracted total Pomeron-proton cross-sections for different cases: a) I; b) II; c) III; d) IV at $t=-0.05\,\mbox{GeV}^2$. Related semi-quantitative description of the data is depicted in Figs.~\ref{fig:dsigSDCDFII}-\ref{fig:dsigSDTOTEMt}. Solid curves represent the data at $\sqrt{s}=546$~GeV, dashed curves are related to $\sqrt{s}=1800$~GeV, and dash-dotted are related to $\sqrt{s}=7$~TeV.}
\end{figure}

\begin{figure}[hbt!]  
 \includegraphics[width=0.22\textwidth]{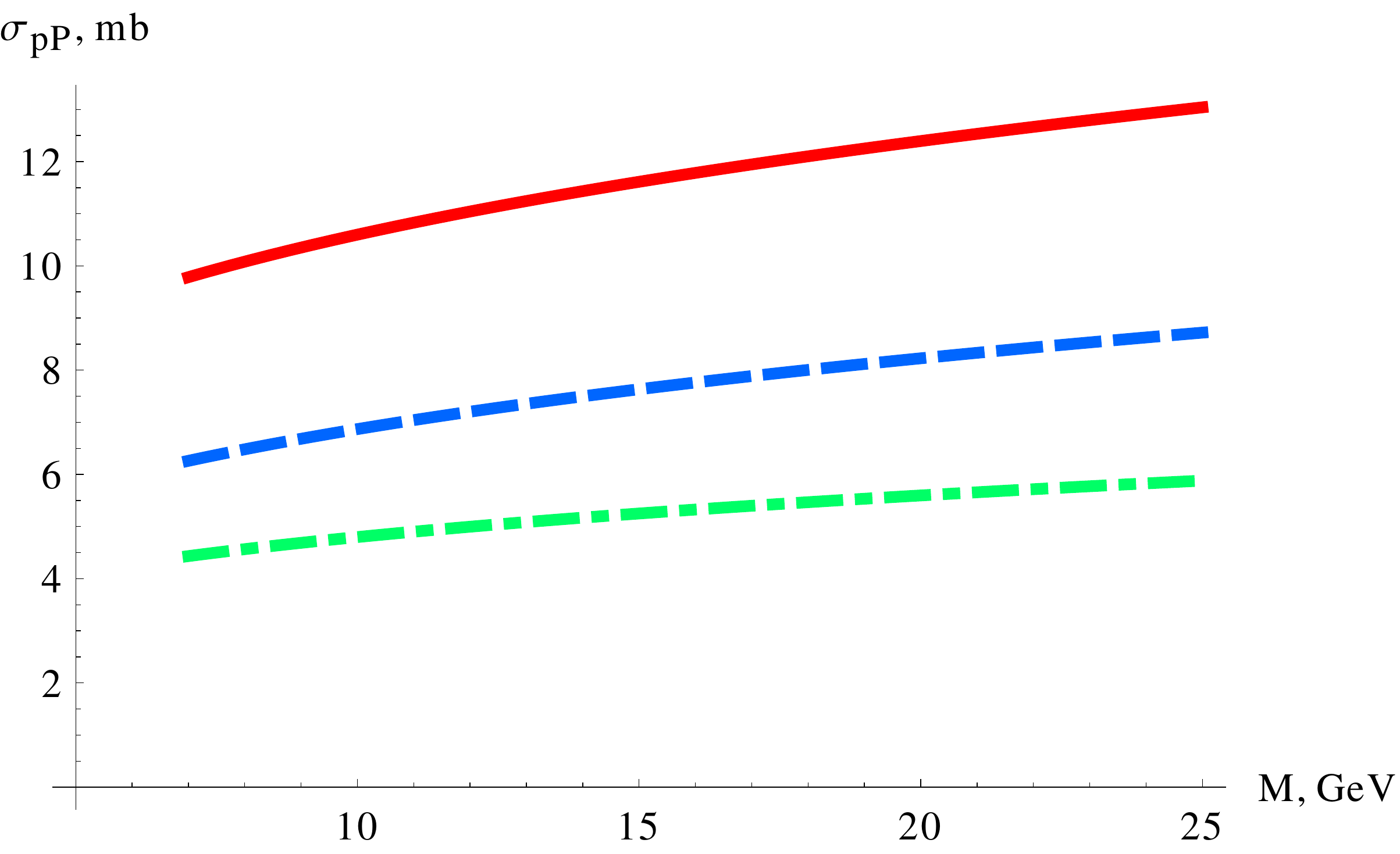} 
 \includegraphics[width=0.22\textwidth]{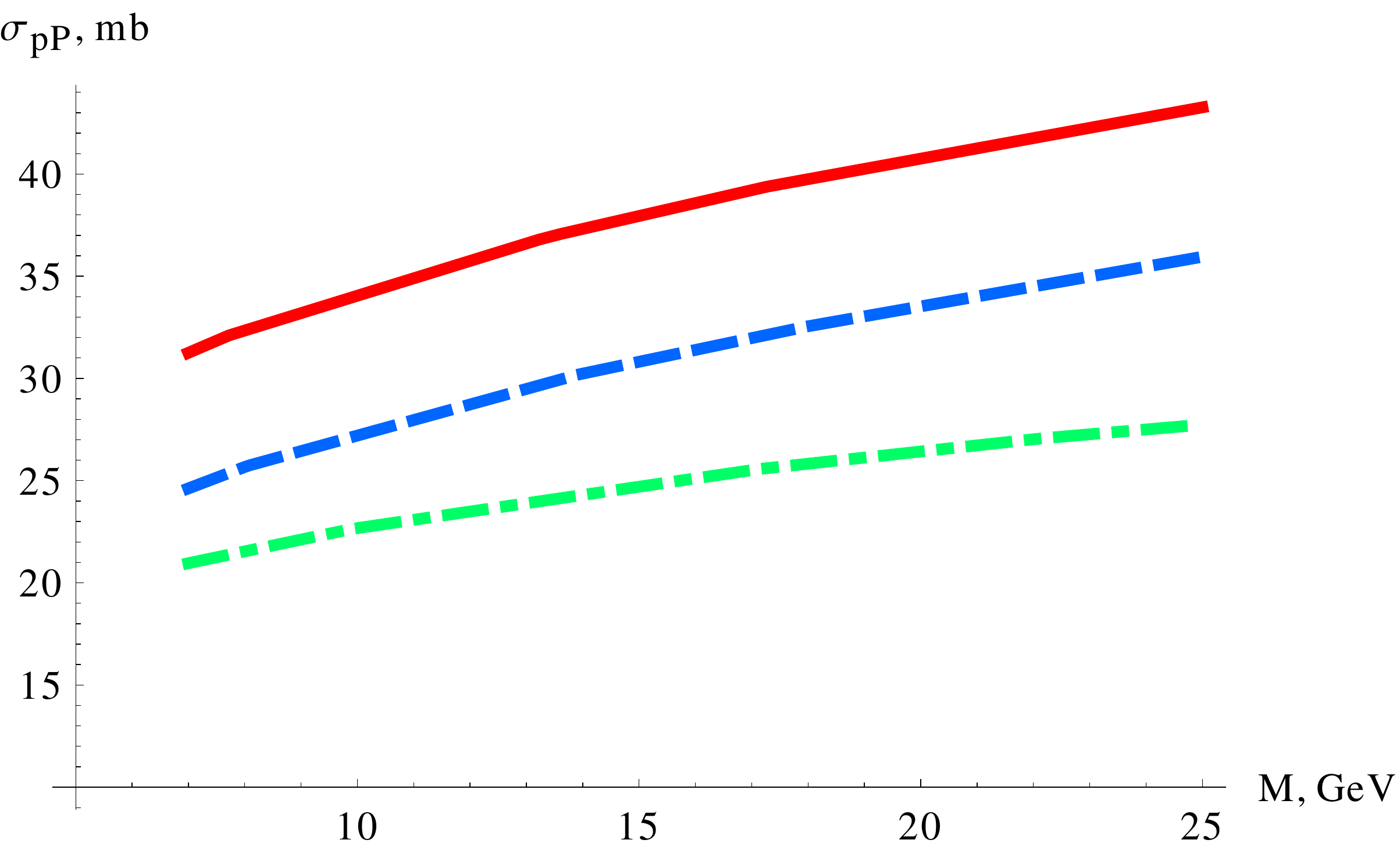}\\
 \hspace*{0.11\textwidth}a)\hspace*{0.20\textwidth}b)\\
 \includegraphics[width=0.22\textwidth]{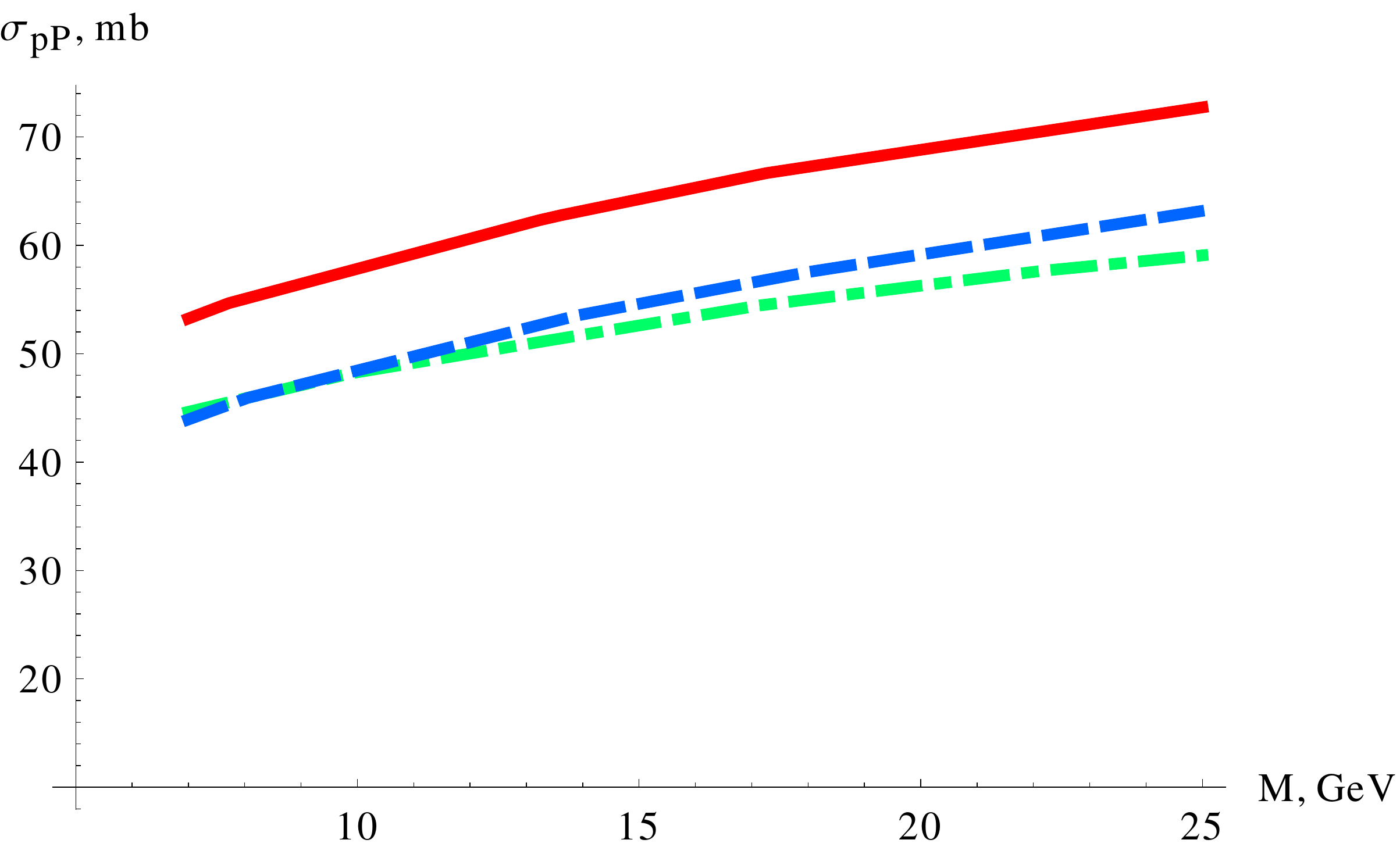}
 \includegraphics[width=0.22\textwidth]{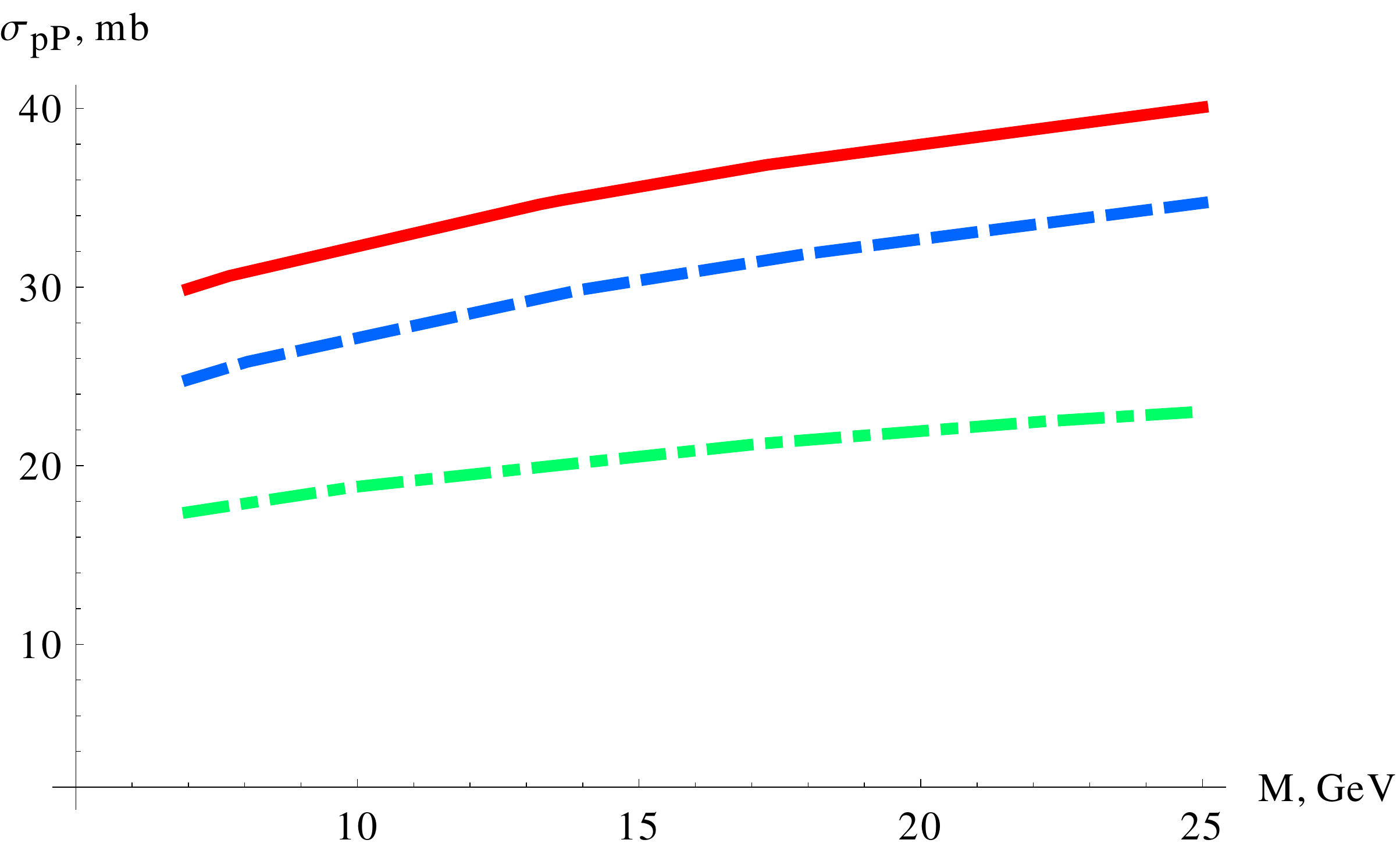}\\
 \hspace*{0.11\textwidth}c)\hspace*{0.20\textwidth}d)
  \caption{\label{fig:sigpPM3} M-dependence of the extracted total Pomeron-proton cross-sections for different cases: a) I; b) II; c) III; d) IV at $t=-0.1\,\mbox{GeV}^2$. Related semi-quantitative description of the data is depicted in  Figs.~\ref{fig:dsigSDCDFII}-\ref{fig:dsigSDTOTEMt}. Solid curves represent the data at $\sqrt{s}=546$~GeV, dashed curves are related to $\sqrt{s}=1800$~GeV, and dash-dotted are related to $\sqrt{s}=7$~TeV.}
\end{figure}

Let us consider other cases. As we can see from 
Figs.~\ref{fig:sigpPt}-\ref{fig:sigpPM3}(b) (the case II), when 
we suppose $t$-independent pro\-ton-Po\-me\-ron
cross-section, the unitarization does not change the situation much. But in cases 
III, IV (Figs.~\ref{fig:sigpPt}-\ref{fig:sigpPM3}(c),(d)) with $t$-dependent proton-Po\-me\-ron
cross-sec\-tion the situation is better when we take into account only the initial state 
rescattering.

\item As to the experimental data on SD, it is not so numerous in the kinematic
region of the Po\-me\-ron dominance~(\ref{eq:kinmod1}). Since that we use it
only for a semi-quan\-ti\-ta\-tive analysis, which is depicted 
in Figs.~\ref{fig:dsigSDCDFII}-\ref{fig:dsigSDTOTEMt}. More or less appropriate
data can be found in~\cite{data1}-\cite{data4}. To illustrate some aspects of
the data analysis we use fits with errors (depicted 
as filled areas in figures).

\begin{figure}[hbt!]  
 \includegraphics[width=0.22\textwidth]{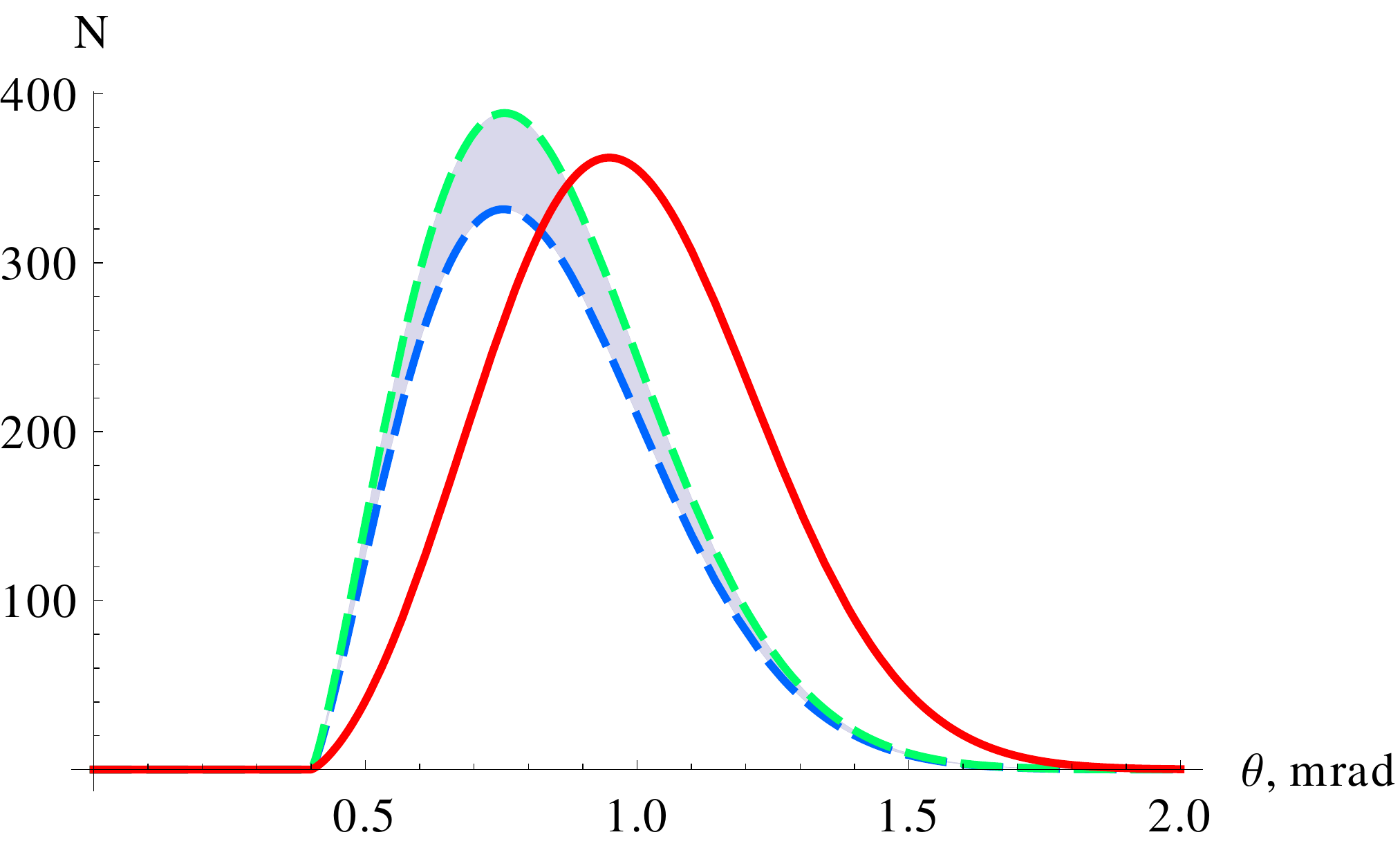} 
 \includegraphics[width=0.22\textwidth]{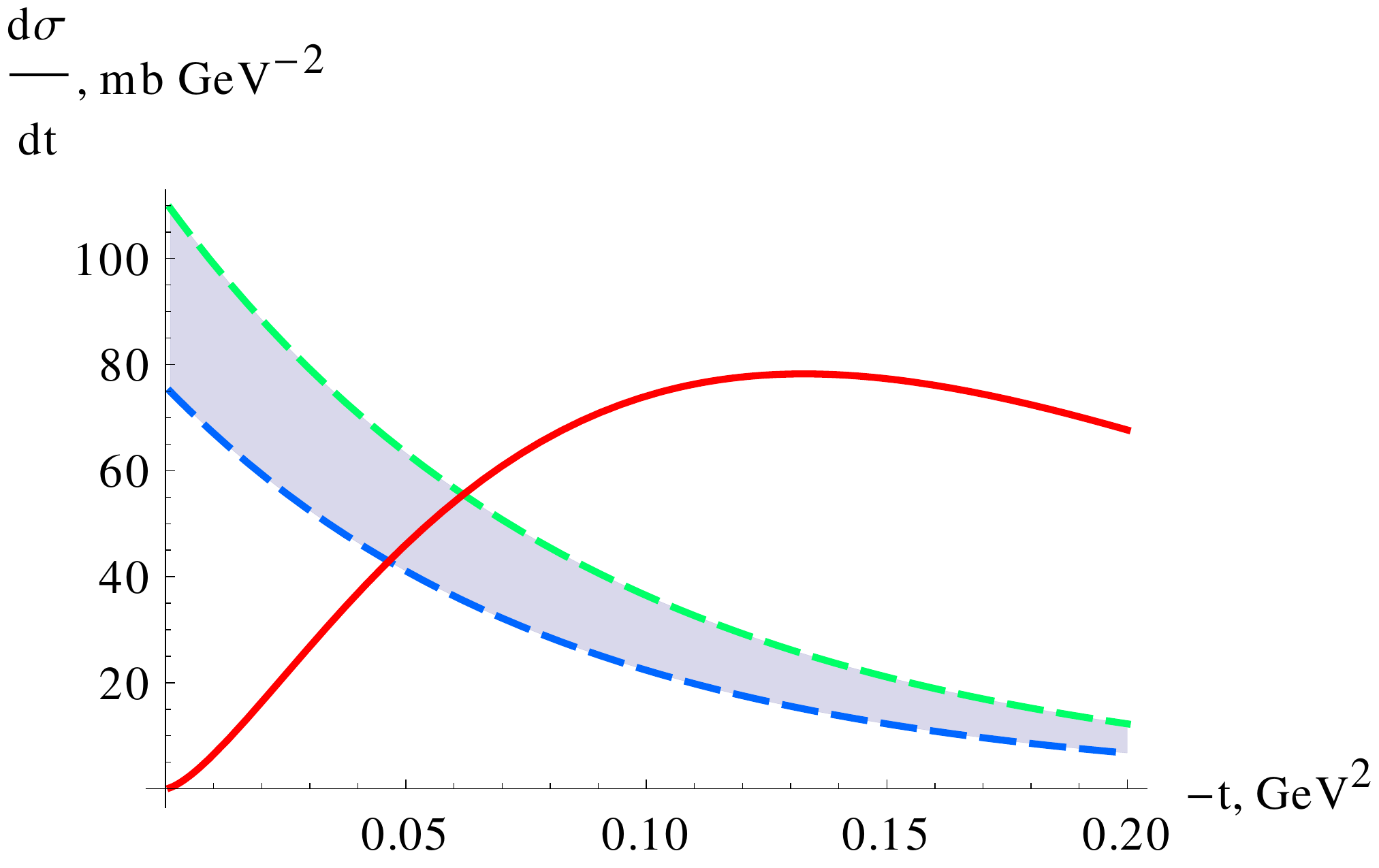}\\
 \hspace*{0.11\textwidth}a)\hspace*{0.20\textwidth}b)\\
 \includegraphics[width=0.22\textwidth]{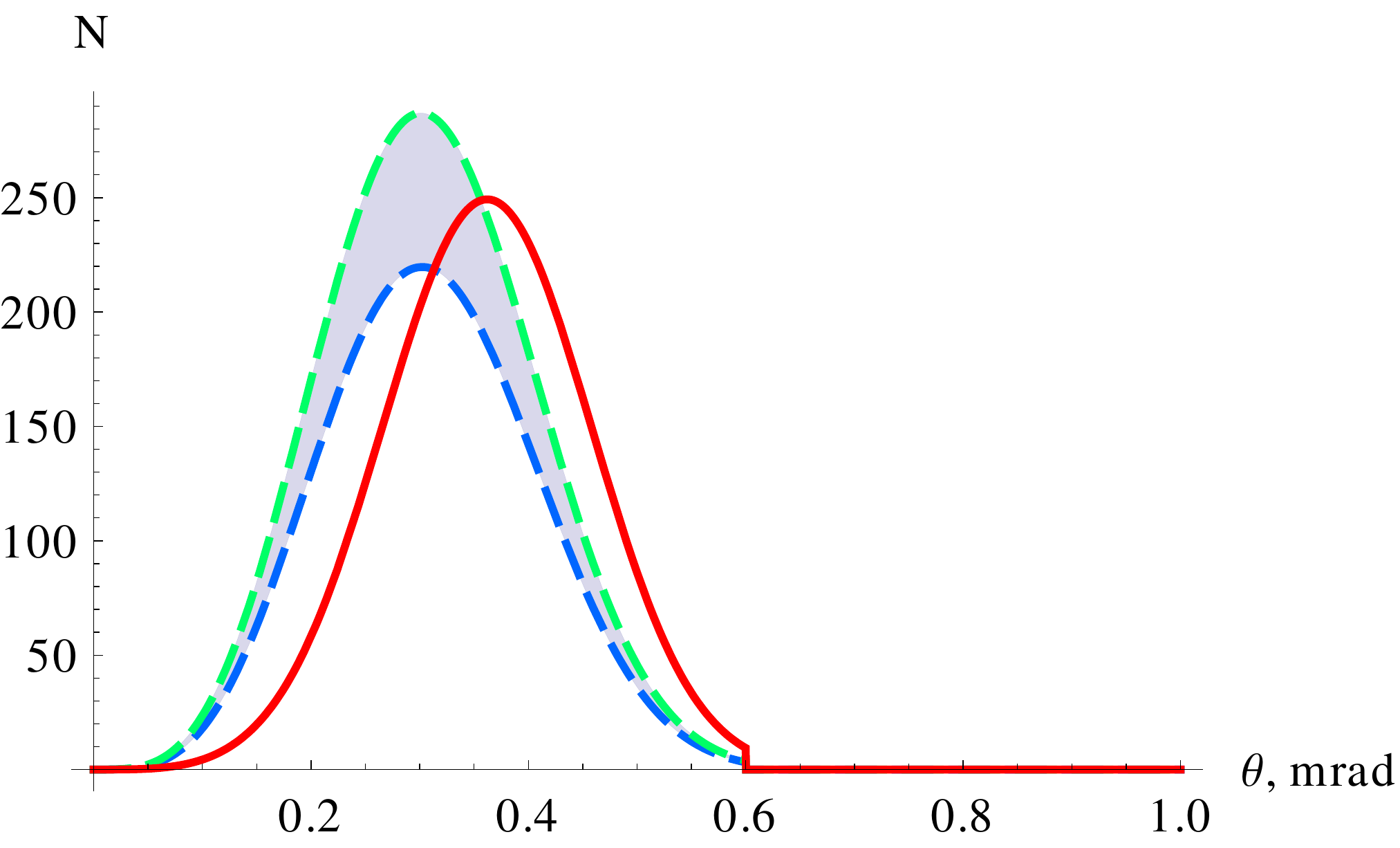}
 \includegraphics[width=0.22\textwidth]{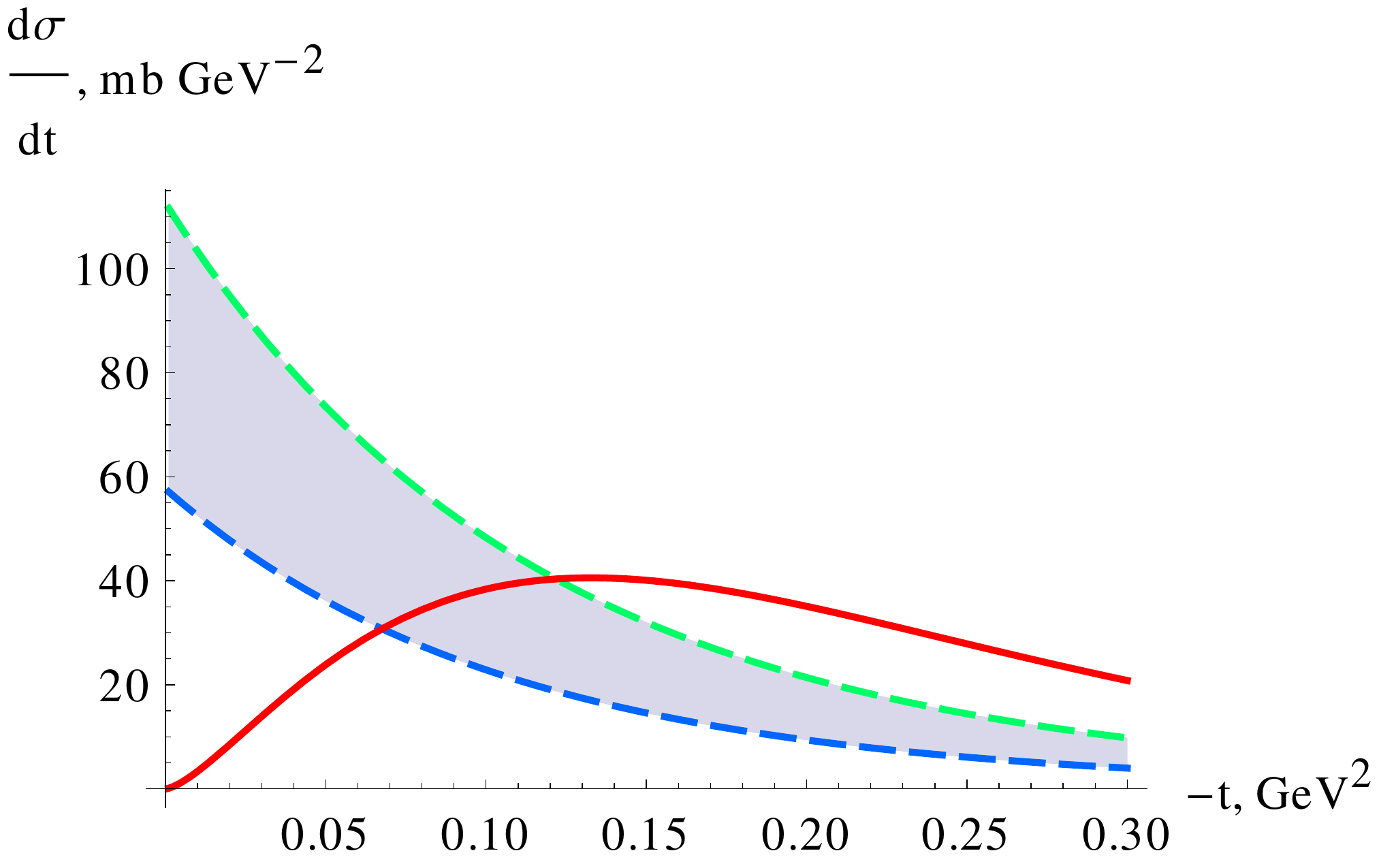}\\
 \hspace*{0.11\textwidth}c)\hspace*{0.20\textwidth}d)
  \caption{\label{fig:dsigSDCDFII} Semi-quantitative description (solid curves) of the CDF data at $\sqrt{s}=546$~GeV (pictures (a), (b)) and $\sqrt{s}=1800$~GeV (pictures (c), (d)) for the case II. In (a), (c) filled areas between dashed curves approximately represent figures 13b, 14b of Ref.~\cite{data1} ($N$ versus $\theta$). In (b), (d) the same curves represent the fit (3) of Ref.~\cite{data1} ($d\sigma_{\mbox{{\tiny SD}}}/dt$ versus $t$).}
\end{figure}

\begin{figure}[hbt!]  
 \includegraphics[width=0.22\textwidth]{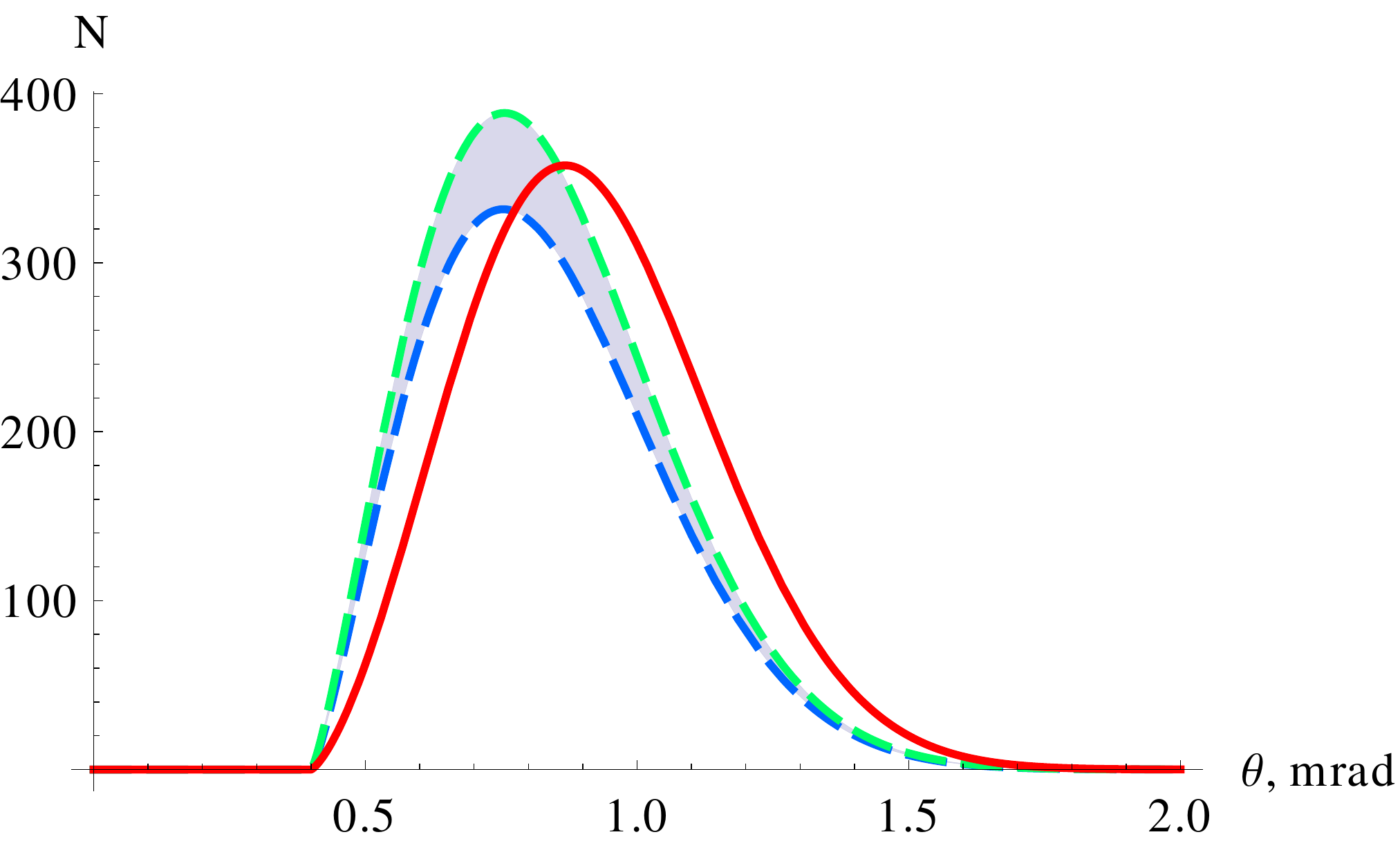} 
 \includegraphics[width=0.22\textwidth]{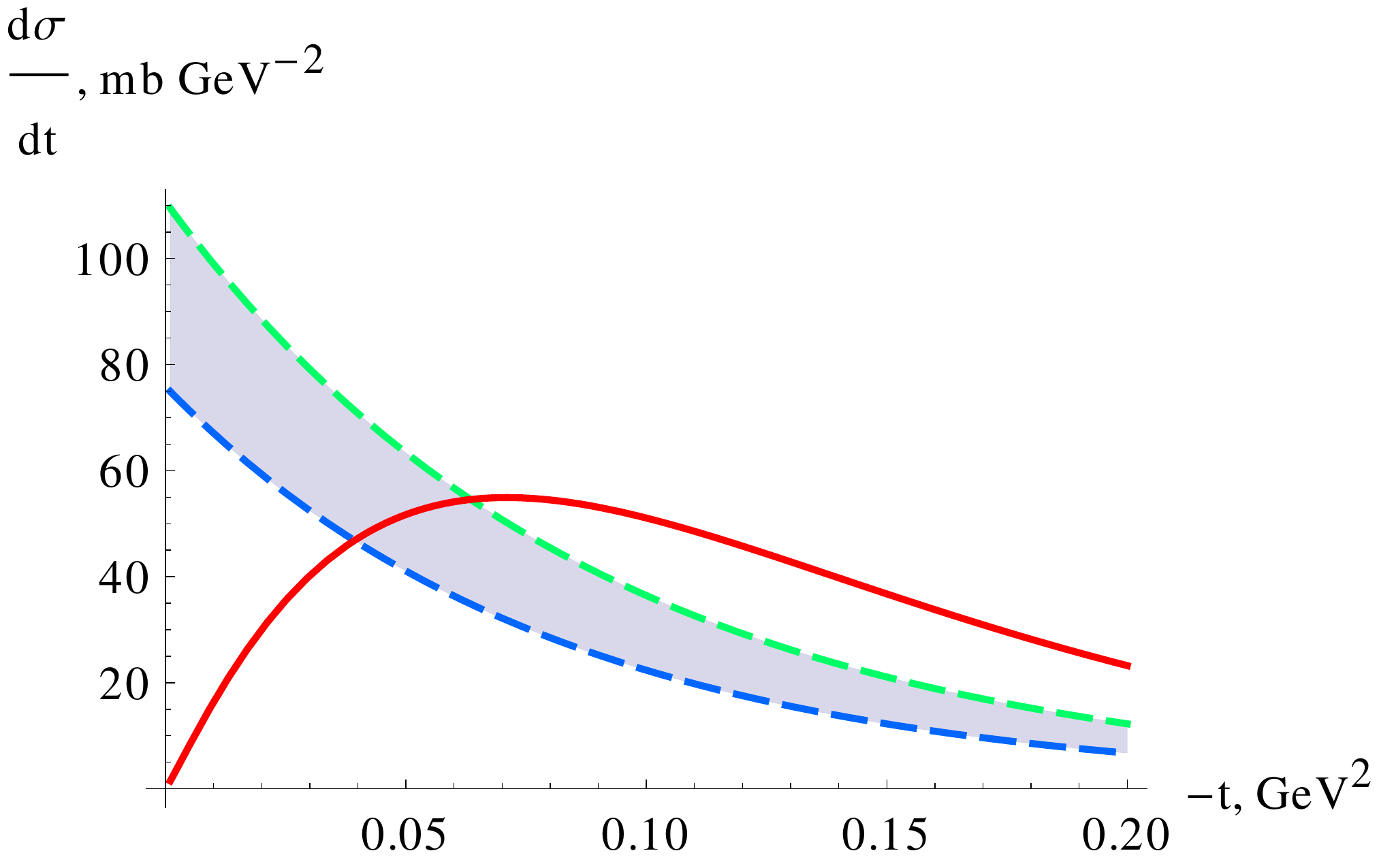}\\
 \hspace*{0.11\textwidth}a)\hspace*{0.20\textwidth}b)\\
 \includegraphics[width=0.22\textwidth]{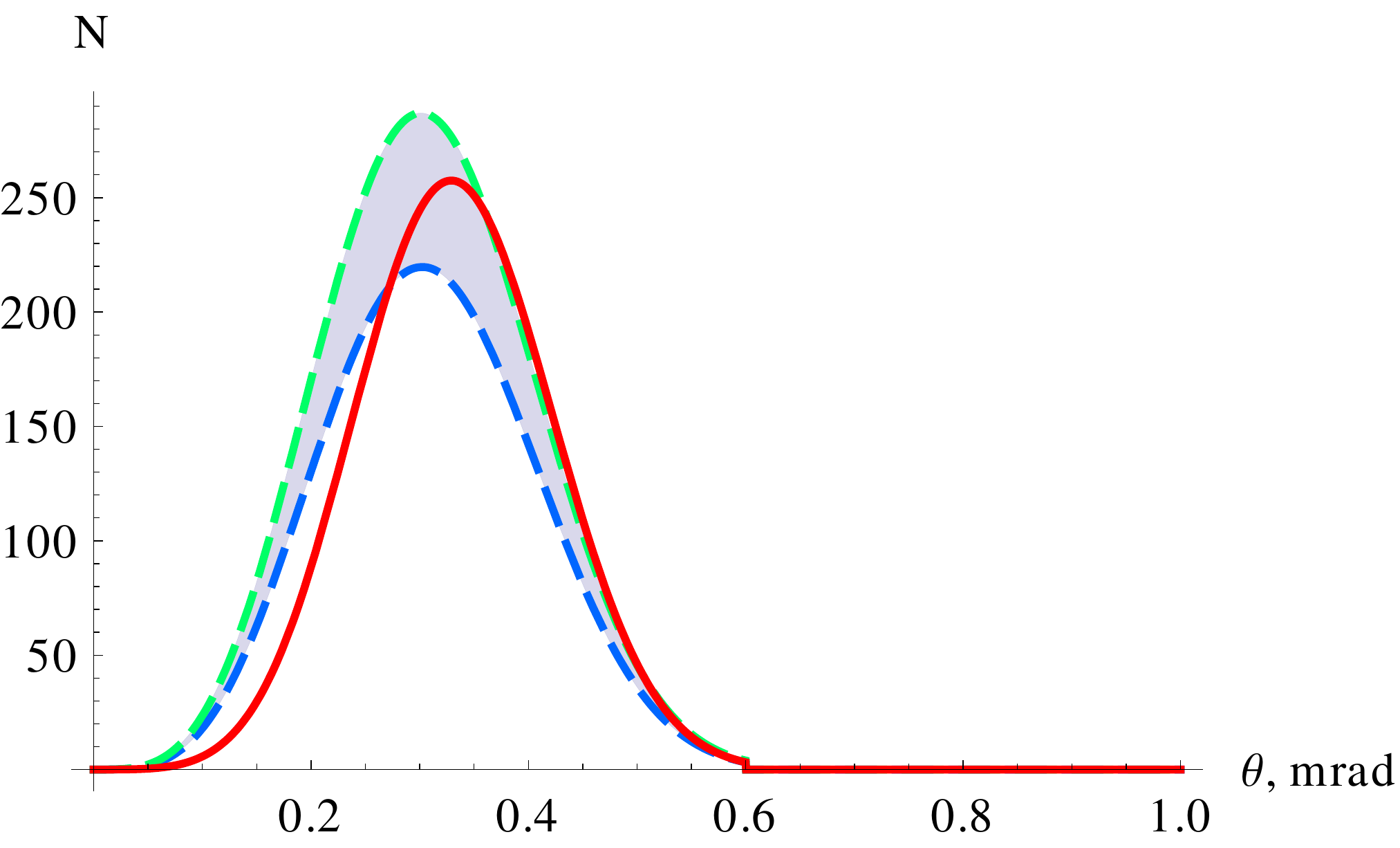}
 \includegraphics[width=0.22\textwidth]{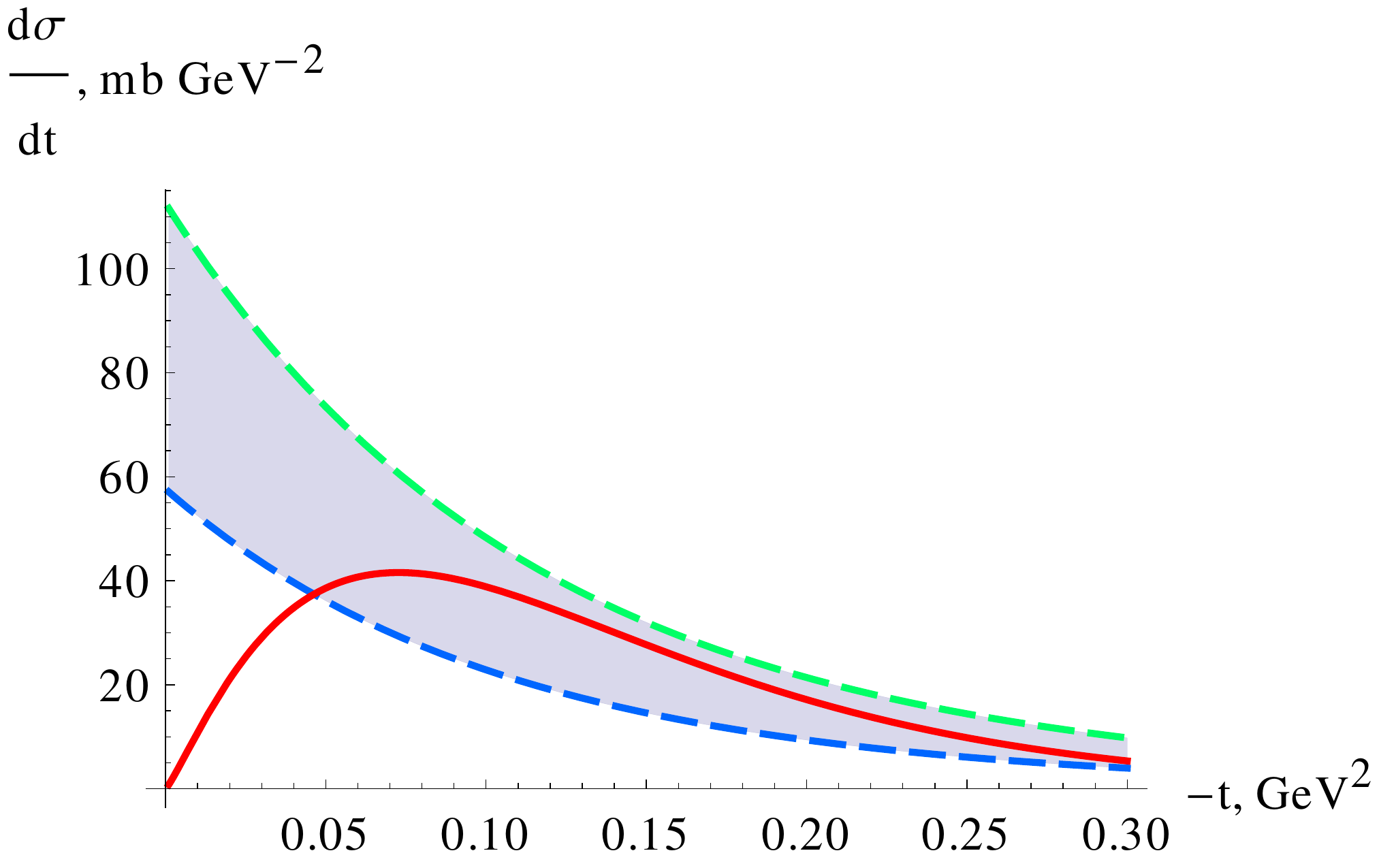}\\
 \hspace*{0.11\textwidth}c)\hspace*{0.20\textwidth}d)
  \caption{\label{fig:dsigSDCDFIII} Semi-quantitative description (solid curves) of the CDF data at $\sqrt{s}=546$~GeV (pictures (a), (b)) and $\sqrt{s}=1800$~GeV (pictures (c), (d)) for the case III. In (a), (c) filled areas between dashed curves approximately represent figures 13b, 14b of Ref.~\cite{data1} ($N$ versus $\theta$). In (b), (d) the same curves represent the fit (3) of Ref.~\cite{data1} ($d\sigma_{\mbox{{\tiny SD}}}/dt$ versus $t$).}
\end{figure}

\begin{figure}[hbt!]  
 \includegraphics[width=0.22\textwidth]{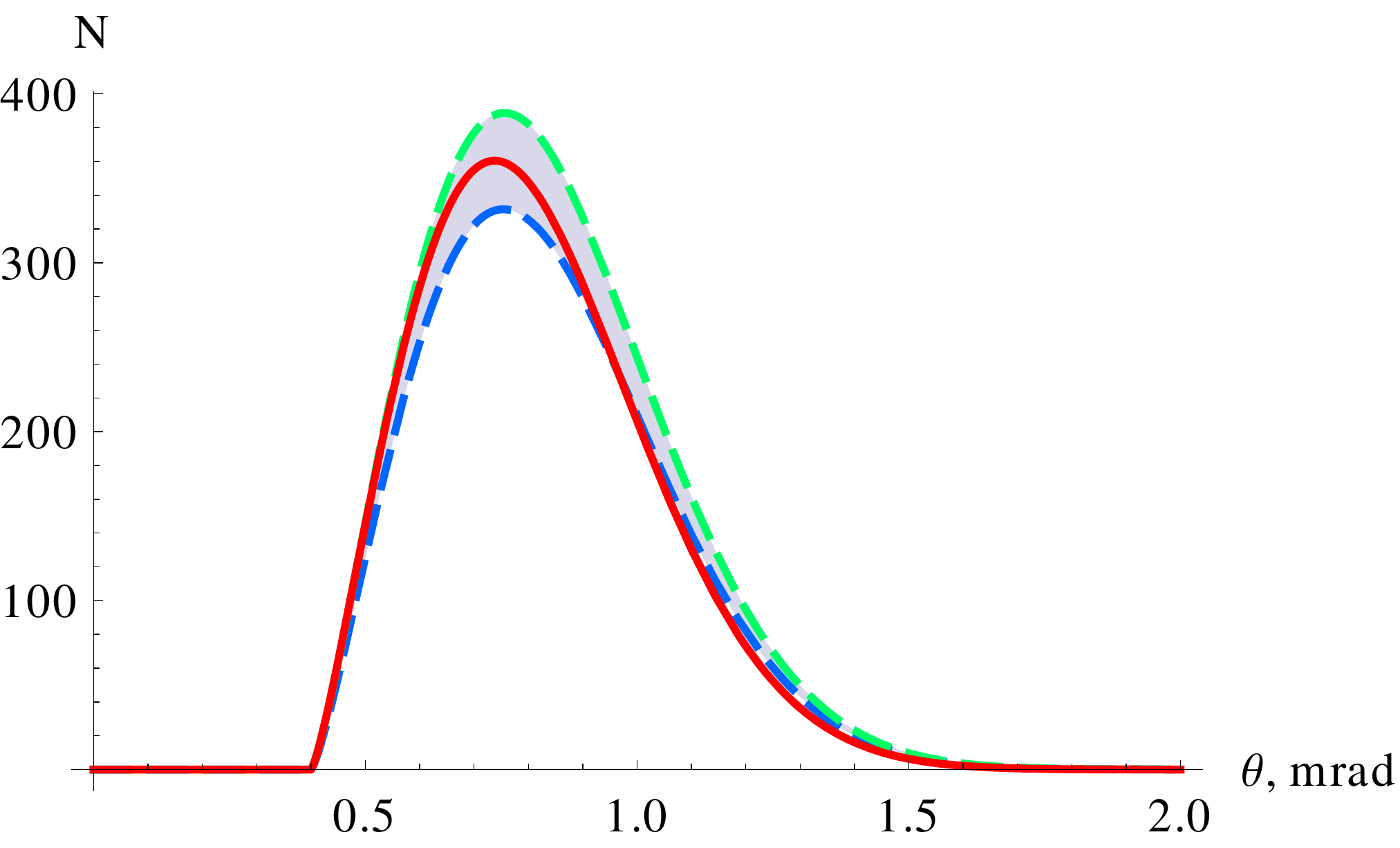} 
 \includegraphics[width=0.22\textwidth]{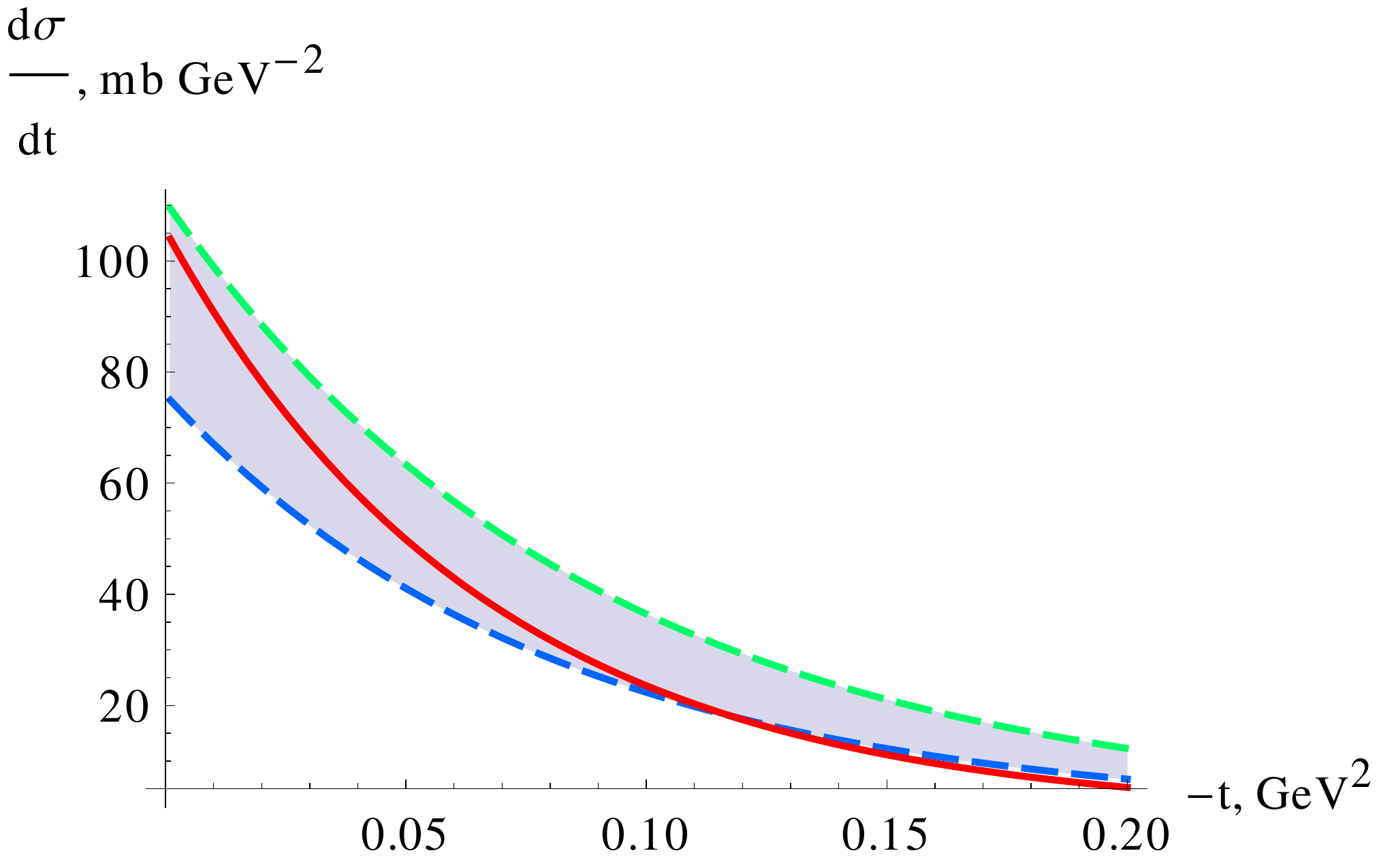}\\
 \hspace*{0.11\textwidth}a)\hspace*{0.20\textwidth}b)\\
 \includegraphics[width=0.22\textwidth]{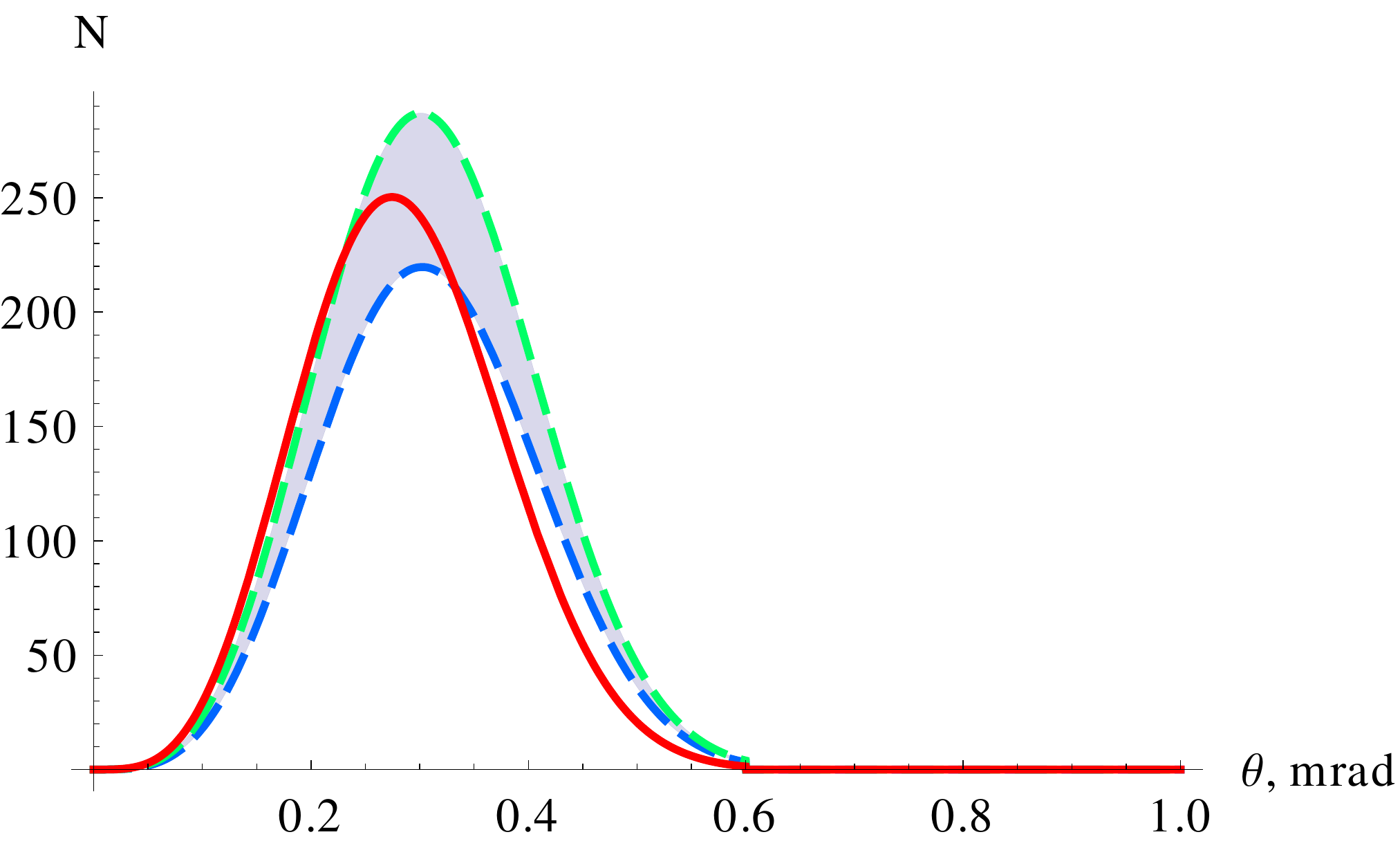}
 \includegraphics[width=0.22\textwidth]{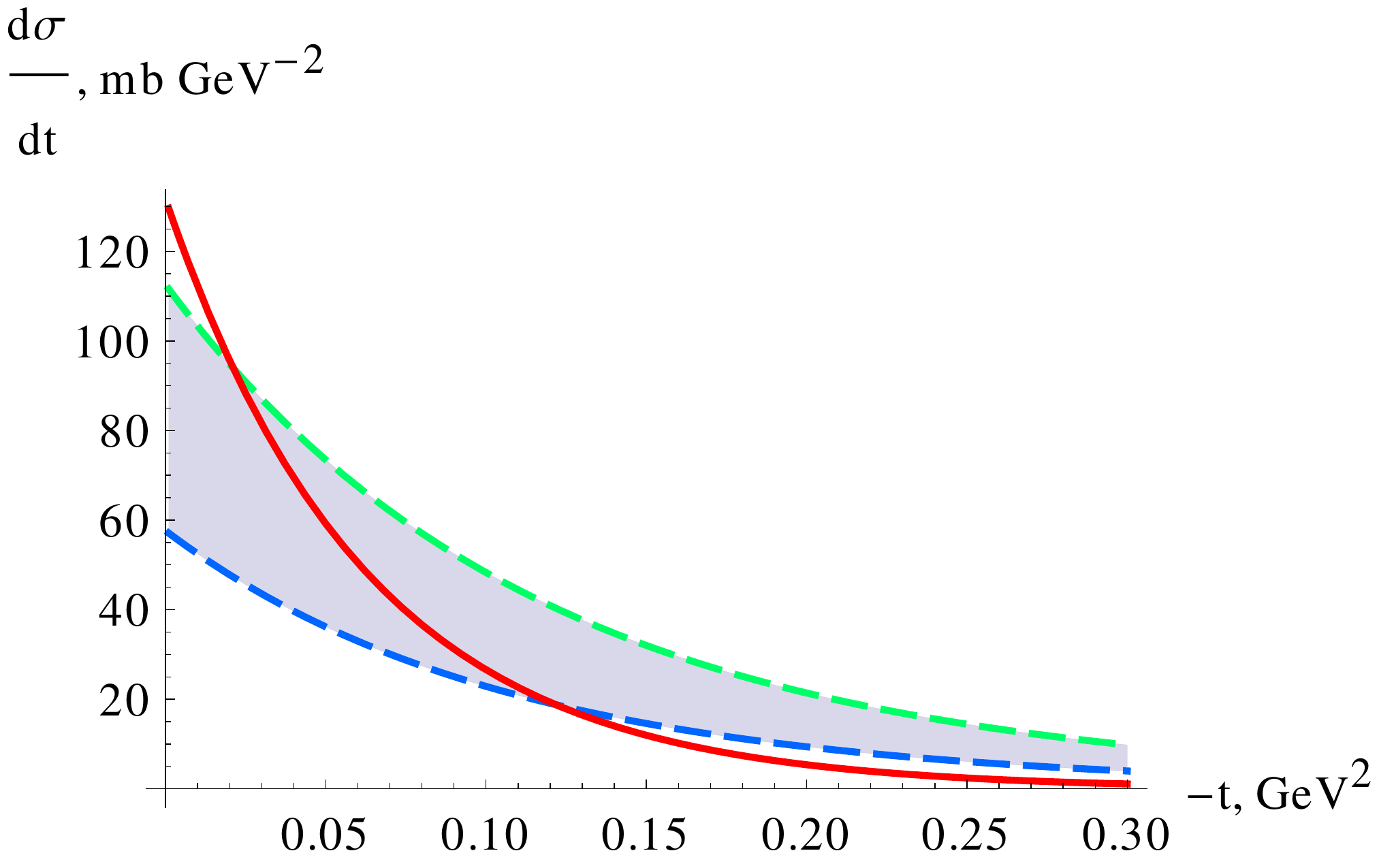}\\
 \hspace*{0.11\textwidth}c)\hspace*{0.20\textwidth}d)
  \caption{\label{fig:dsigSDCDFIV} Semi-quantitative description (solid curves) of the CDF data at $\sqrt{s}=546$~GeV (pictures (a), (b)) and $\sqrt{s}=1800$~GeV (pictures (c), (d)) for the case IV. In (a), (c) filled areas between dashed curves approximately represent figures 13b, 14b of Ref.~\cite{data1} ($N$ versus $\theta$). In (b), (d) the same curves represent the fit (3) of Ref.~\cite{data1} ($d\sigma_{\mbox{{\tiny SD}}}/dt$ versus $t$).}
\end{figure}

\begin{figure}[hbt!]  
 \includegraphics[width=0.22\textwidth]{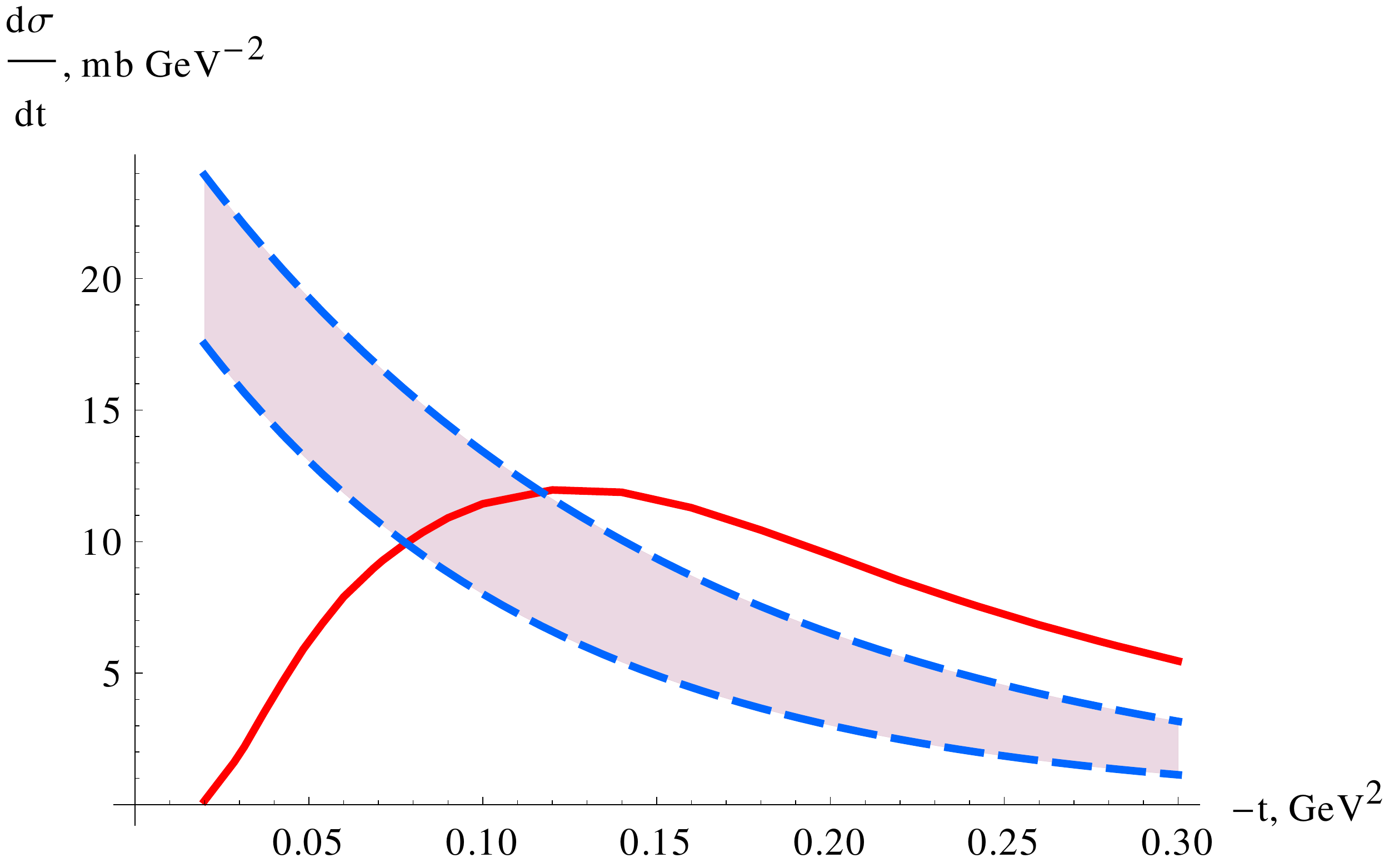} 
 \includegraphics[width=0.22\textwidth]{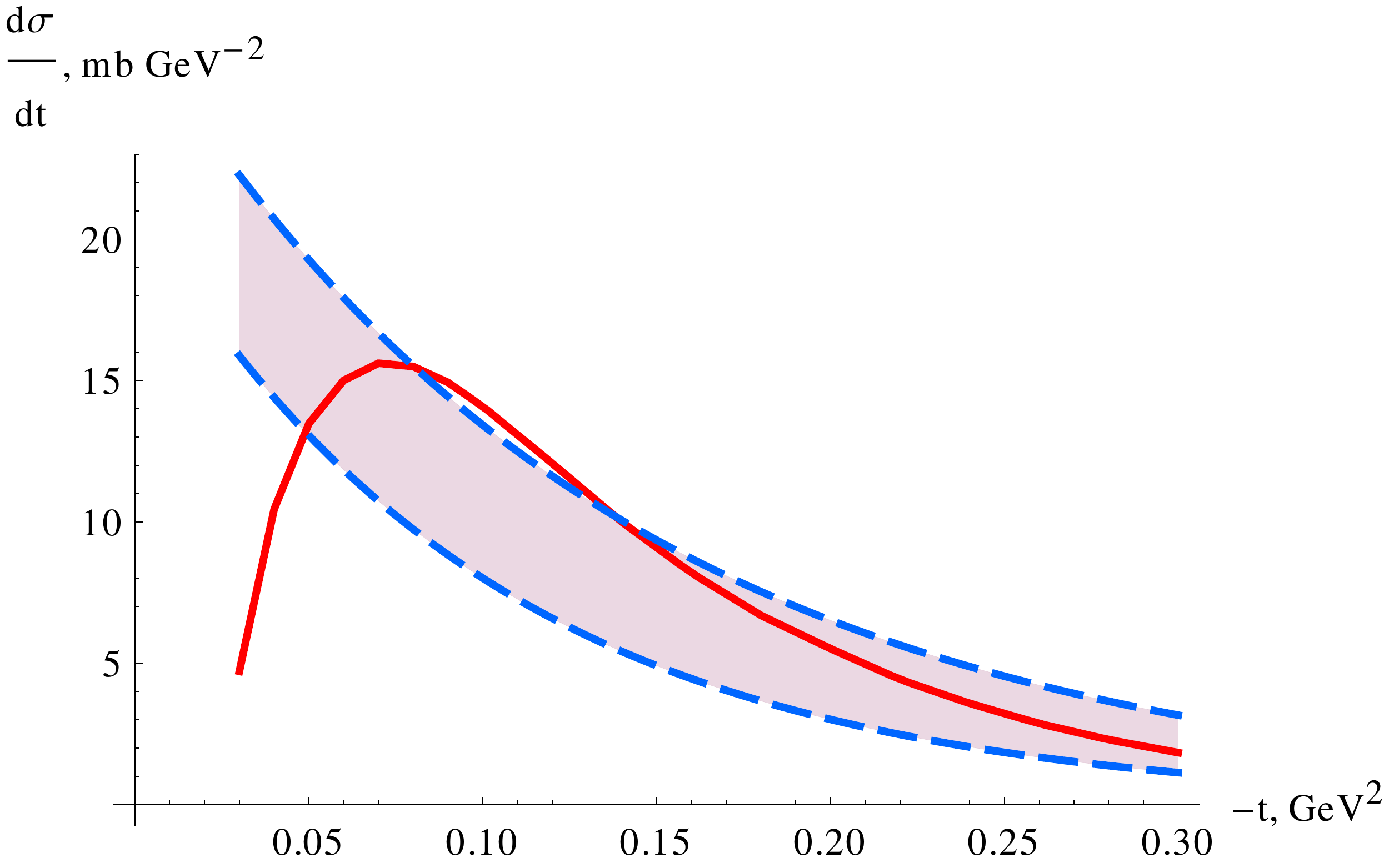}\\
 \hspace*{0.11\textwidth}a)\hspace*{0.20\textwidth}b)\\
 \includegraphics[width=0.22\textwidth]{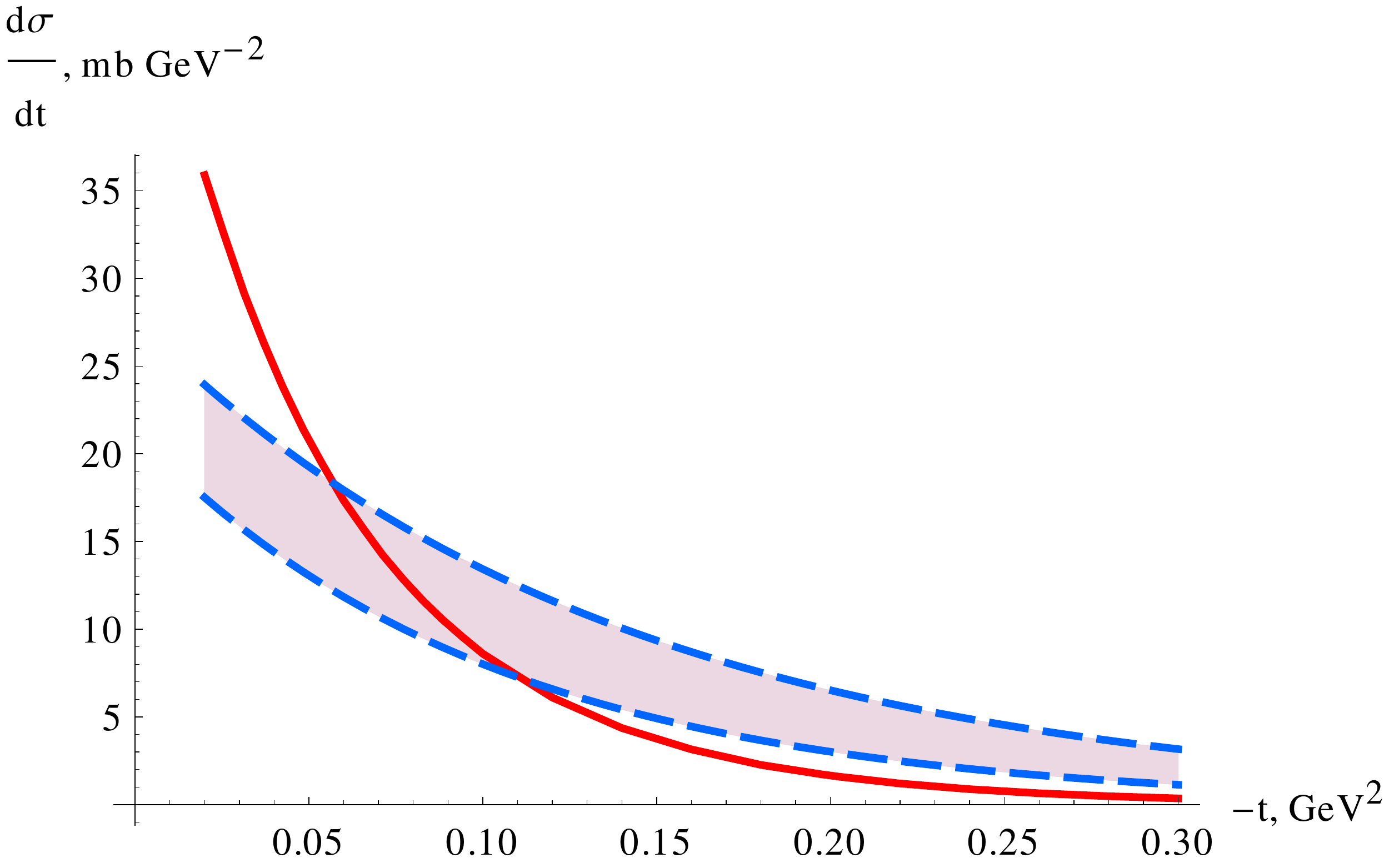}\\
 \hspace*{0.11\textwidth}c)
  \caption{\label{fig:dsigSDTOTEMt} Semi-quantitative description (solid curves) of the TOTEM data at $\sqrt{s}=7$~TeV~\cite{data3}-\cite{data4} for cases: a) II; b) III; c) IV. Filled areas between dashed curves approximately represent the TOTEM data on $d\sigma_{\mbox{{\tiny SD}}}/dt$ in the mass region from $7$ up to $350$~GeV. We do not consider here the case I, since the need of rescattering corrections is clear from next figures.}
\end{figure}

 In Figs.~\ref{fig:dsigSDCDFII}-\ref{fig:dsigSDTOTEMt} we consider only cases II-IV, since the need of
 unitarization is obvious from the case I. First, we show the CDF data~\cite{data1} as a function of $\theta$ variable
 in pictures~\ref{fig:dsigSDCDFII}-\ref{fig:dsigSDCDFIV}(a),(c). The reason is that
 the function of the acceptance can cut out the region of the special maximum
 of the differential cross-section, and the data can be succesfully fitted by our
 curves (in the model of conserved tensor currents). This becomes clear, for example, from the
 Fig.~\ref{fig:dsigSDCDFIII}(c),(d), where we have rather good description of the data at $\sqrt{s}=1800$~GeV even
 if the situation with $t$-dependence is worse. Let us note that this case (III) is the most consistent, since
for getting the $t$-dependence  of the proton-Po\-me\-ron cross-section  we can use the 3-Pomeron vertex. More singular behaviour of the case
 IV looks rather odd.

 We can conclude from the above analysis that in the current conservation approach the proton-Po\-me\-ron
 cross-section should be essentially $t$-dependent to fit accurately the existing data on SD, including also
 the latest experimental results from the 
 TOTEM~\cite{data3},\cite{data4}. As to the
 data from other collaborations~\cite{data5}-\cite{data8}, they present only integrated cross-sections and sometimes without
 the proton tagging (with rapidity gaps only). It 
 is difficult to extract the proton-Po\-me\-ron cross-section for this case in a more or less
 model independent way since we do not 
 know its exact behaviour in $t$ .
 
\end{itemize}

\section*{Conclusions}

This paper is an attempt to highlight 
some important aspects of 
the present theoretical and 
experimental situation in SD 
and DD. We have used the method of the
covariant reggeization with conserved tensor 
currents. In this special approach 
basic conclusions can be summarized
in the following list:

\begin{itemize}
\item Rescattering corrections should be taken into account in an appropriate way, since
the case I shows s-de\-pen\-dent proton-Po\-me\-ron cross-section, which looks 
as a nonsense.
\item Even if we take into account unitarization(in the form available to us at present), the 
total proton-Po\-me\-ron
cross-section should depend somehow on $t$ to be independent on $s$. This is obvious
from the case II, when the cross-section is $t$-independent.
\item The data on SD is not so numerous in the kinematical
region of the Pomeron dominance~(\ref{eq:kinmod1}). The basic conclusion
after the analysis is the following: more singular $t$-dependence (until 
the complete cancellation of the special dependence at small $t$ in the case IV)
in the proton-Po\-me\-ron cross-section leads to better 
description of the data. Are these data reliable enough or not? At this moment
we are not sure about it. We hope that further LHC experiments will
improve the situation.
\item Finally, the question of tensor current conservation that leads to
rather singular $t$-dependence in the proton-Po\-me\-ron cross-section, can
be resolved, if we have good quality data on SD and DD differential 
cross-sections.

Our next subject will be the case of nonconserved
currents (which factually is usually being used in literature) 
as an alternative view of SD and 
DD dynamics.
\end{itemize}

We are to emphasize again that from the theory side the most difficult problems 
is to find a proper unitarization scheme.
 
\section*{Acknowledgements}

We are grateful to Anton Godizov for useful discussions.

\end{document}